\documentclass[12pt, twocolumn]{aastex61}

\usepackage{multirow}
\usepackage{ulem}
\newcommand{\gf}{ }

\renewcommand{\vec}[1]{ {\mathbf #1} }

\def\mw{{microwave}}

\received{July 1, 2016}
\revised{September 27, 2016}
\accepted{\today}
\submitjournal{ApJ}

\shorttitle{NLFFF Reconstructions with Chromospheric Magnetic Field Data}
\shortauthors{MHD/NLFFF Team}
\begin{document}


\title{Force-Free Field Reconstructions Enhanced by Chromospheric Magnetic Field Data}
%{ Adding Chromospheric Magnetic Field Data to the Coronal Magnetic Field Reconstruction Tools}
%Casting the Coronal Magnetic Field Reconstruction Tools  in 3D Using MHD Bifrost Model
\author{Gregory Fleishman$^1$}
%\affil{Physics Department, Center for Solar-Terrestrial Research, New Jersey Institute of Technology
%Newark, NJ, 07102-1982}

\author{Ivan Mysh'yakov$^2$}
%\affil{Institute of Solar-Terrestrial Physics SB RAS}
%
% 	
\author{Alexey Stupishin$^3$}
%\affil{Saint Petersburg State University}
%
%
%
\author{Maria Loukitcheva$^{3,4}$}
%\affil{New Jersey Institute of Technology}
%
%\affil{Saint Petersburg State University}
\author{Sergey Anfinogentov$^2$}
%\affil{Institute of Solar-Terrestrial Physics SB RAS}

%\author{Gregory Fleishman\altaffiliation{1}, Sergey Anfinogentov\altaffiliation{2}, Maria Loukitcheva\altaffiliation{1,3}, Ivan Mysh'yakov\altaffiliation{2}, Alexey Stupishin\altaffiliation{3}}
%\vspace{1cm}
\altaffiliation{$^1$Physics Department, Center for Solar-Terrestrial Research, New Jersey Institute of Technology
Newark, NJ, 07102-1982}

\altaffiliation{$^2$Institute of Solar-Terrestrial Physics  (ISZF), Lermontov st., 126a, Irkutsk, 664033  Russia}

\altaffiliation{$^3$Saint Petersburg State University, 7/9 Universitetskaya nab., St.
Petersburg, 199034 Russia}

\altaffiliation{$^4$Saint Petersburg branch of Special Astrophysical Observatory, Pulkovskoye chaussee 65/1, St. Petersburg 196140, Russia}

%\altaffiltext{4}{Visiting Programmer, Space Telescope Science}

\begin{abstract}

3D picture of the coronal magnetic field remains an outstanding problem in solar physics, particularly, in active regions.
Nonlinear force-free field reconstructions that employ routinely available full-disk photospheric vector magnetograms represent  state-of-the-art coronal magnetic field modeling.
Such reconstructions, however, suffer from an inconsistency between a force-free coronal magnetic field and non-force-free photospheric boundary condition, from which the coronal reconstruction is performed. %Realistic time-dependent full-fledged MHD modeling could help greatly, but this is not expected to be routinely availble in a foreseeable future.
%The use of chromospheric vector magnetograms can aid the coronal part of the magnetic {field} model, but does not help to build the magnetic {field} model between the photospheric and chromospheric levels.
%However, adding some chromospheric magnetic field data on top of available photospheric vector magnetograms is already possible now and the amount and quality of the chromospheric and coronal magnetic diagnostics is supposed to grow steadily.
In this study we focus on integrating the additional chromospheric and / or coronal magnetic field data with the vector photospheric magnetograms with the goal of improving the reliability of the  magnetic field reconstructions. We develop a corresponding modification of the available optimization codes described in Fleishman et al. (2017) and test their performance using a full-fledged MHD model obtained from the Bifrost code by performing a ``voxel-by-voxel'' comparison between the reconstructed and the model magnetic fields.  We demonstrate that adding even an incomplete set of chromospheric magnetic field data can measurably improve the reconstruction of the coronal magnetic field,  greatly improve reconstructions of the magnetic connectivity and of the coronal electric current.
\end{abstract}


W\section{Introduction}

% \begin{itemize}
%   \item State the problem of coronal field reconstruction.
%   \item Justify  the need of the reconstruction method validation.
%   \item Describe the value of using realistic MHD models for this validation.
% \end{itemize}

Quantification of the coronal magnetic field remains a central problem in solar physics.
Nowadays, {the} most common approach to the coronal magnetic field reconstruction is based on \textbf{a} static nonlinear force-free field (NLFFF) {model } \citep[see reviews by][]{1989SSRv...51...11S, 1997SoPh..174..129A, 2008JGRA..113.3S02W, 2012LRSP....9....5W, 2014A&ARv..22...78W}. {NLFFF extrapolations are} routinely possible {thanks to} almost uninterrupted availability of the photospheric full-disk vector magnetic field data starting {from the} launch of the \textit{Solar Dynamics Observatory} \citep[\textit{SDO}, ][]{scherrer_et_al_2012, schou_et_al_2012}  in 2010. A more advanced approach based on the dynamic ``data-driven'' modeling with a number of promising advantages over the static one was also attempted \citep[see, e.g.,][]{2016NatCo...711522J, 2018ApJ...852...82Y} but it is not yet widely used as it requires a lot more resources to reconstruct the field than the NLFFF approach.

The NLFFF methods represent a significant advance compared with earlier, more simplistic methods of potential or linear force-free field extrapolations; however, even the much improved and more sophisticated NLFFF reconstruction methods still suffer from a number of known shortcomings \citep[see, e.g.,][]{2009ApJ...696.1780D, 2015ApJ...811..107D}.
One of them, which we focus on in this study, is an inconsistency between the coronal field force-freeness, and the bottom (photospheric) boundary condition, which can significantly deviate from the force-free state.
In our recent study \citep[][hereafter Paper I]{2017ApJ...839...30F}, we used a publicly available dataset \citep[\textit{en024048{\_}hion},][http://sdc.uio.no/search/simulations]{2016A&A...585A...4C} {produced with} the full-fledged  3D radiation magnetohydrodynamic (RMHD) code Bifrost \citep{2011A&A...531A.154G} to cast the coronal magnetic field reconstruction tools. We found  (perhaps, not surprisingly) that
extrapolations performed starting from a chromospheric vector boundary condition yield measurably better results than those starting from the photospheric boundary. Thus, for the coronal magnetic field reconstruction it would be highly beneficial to perform extrapolations starting from a chromospheric layer.
But this has a severe disadvantage that the magnetic field between the photosphere and corresponding chromospheric boundary remains unspecified, although a magnetic {field} model {between these layers} is needed for many practical purposes, e.g., to build magneto-thermal models of active regions within the GX Simulator framework \citep{Nita_etal_2018}.

In practice, even though the chromospheric vector magnetograms are becoming available from the full Stokes spectropolarimetry in the spectral lines formed in the upper chromosphere \citep[most notably HeI 1083\ nm and CaII 854.2\ nm spectral lines;][]{2003SPIE.4853..194K, 2011SPIE.8148E..09B}, these data cannot yet fully substitute the photospheric data for a number of reasons: (i) there are large time gaps in availability of  the chromospheric magnetograms; (ii) obtaining such magnetograms is computationally expensive and less straightforward than in the case of the photosphere; (iii) the derived magnetic field pertains to variable heights unlike the photospheric case; and (iv)  the magnetic field is on average weaker in the chromosphere than in the photosphere, thus the signal-to-noise ratio is weaker that implies bigger errors in the derived magnetic field. This is especially severe for the transverse component of the magnetic field, which has larger errors than the line-of-sight (LOS) component.

There are other ways of the magnetic field probing at the chromospheric or coronal heights, primarily, from the radio measurements \citep{Lee_2007}. One method uses radio polarimetric measurements of the free-free emission \citep{1980SoPh...67...29B, Grebinskij_etal_2000, 2017arXiv170206018L} to yield the LOS component of the magnetic field at a height, where the free-free emission at the given frequency is formed. In a long run such ``tomographic'' diagnostics could be available (at least for some cases) from multi-frequency observations with {\it Atacama Large Millimeter/Submillimeter Array} (ALMA), which will be capable of providing the $B_{\|}$ component at various heights with high accuracy and spatial resolution \citep{2017arXiv170206018L}. So far, this method has been tested on \mw\ data obtained with {\it Nobeyama Radioheliograph} \citep[NoRH,][]{Nakajima_etal_1994} at a single frequency, 17~GHz, with a modest spatial resolution, which proved the concept but remains insufficient for the coronal magneic field reconstruction.

Another method relies on polarized imaging data of the gyroresonant emission \citep{1997SoPh..174...31W}, which yields the absolute value of the magnetic field at the transition region level from the images directly and can also provide a limited information about the magnetic field direction. This method has been tested with a number of the data sets including the data obtained with {\it Very Large Array} (VLA) and {\it Radio Astronomical Telescope of the Academy of Sciences 600} (RATAN-600) \citep{1982SoPh...79...41A,1986ApJ...301..460A,2012ARep...56..790K}, {\it Owens Valley Solar Array} \citep[OVSA,][]{1994ApJ...420..903G}, NoRH \citep{2011SoPh..273..309S}, combination of OVSA and VLA \citep{2011ApJ...728....1T}, combination of NoRH, RATAN-600, and {\it Siberian Solar Radio Telescope} (SSRT) \citep{Nita_etal_2011} and in some other studies. Provided the \mw\ images are available, the method itself is relatively simple \citep{2015ApJ...805...93W}; however, it only provides the absolute value of the magnetic field rather {than} the magnetic field vector. In principle, having a \mw\ imaging instrument with a broad spectral coverage could supply us with both the gyroresonant and free-free diagnostics from same locations simultaneously, but this anyway yields only two of three vector components.

Therefore, we conclude that in the foreseeable future there will not be routine chromospheric vector magnetograms capable to fully substitute the photospheric vector data in the NLFFF reconstructions. However, we expect a progressively more and more mature chromospheric magnetic field data to become available in addition to the photospheric probing. For example, this can be the vector data from a chromospheric subarea, where the magnetic field is strong enough to be reliably measured using the optical spectropolarimetry technique, or radio diagnostics of the LOS component, or the absolute value, or both. In this paper, we address a question if adding a given chromospheric data set in addition to the full vector photospheric boundary condition can help improving the coronal magnetic field reconstruction and to what extent. Following Paper I, we use the same set of testing data derived from the full-fledged RMHD model \citep{2016A&A...585A...4C} to quantify the potential improvement of the reconstruction with one or another set of additional chromospheric magnetic field data.

%\newpage

\section{RMHD Model Data Cube}
\label{S_dcube_overview}

The RMHD simulation that we use for our tests is described by \citet{2016A&A...585A...4C}, of which we employ snapshot 385 that has already been used to perform various kinds of studies \citep{2013ApJ...772...89L, 2013ApJ...772...90L, 2013ApJ...778..143P, 2015ApJ...806...14P, 2015ApJ...811...80R, 2015ApJ...811...81R, 2015ApJ...813...34L, 2012ApJ...749..136L,2015ApJ...802..136L,2013ApJ...764L..11D,2012ApJ...758L..43S,2015ApJ...803...65S, 2015A&A...575A..15L, 2015ASPC..499..349L, 2017arXiv170206018L, 2017ApJ...839...30F}. In Paper I, we interpolated  magnetic field components  from the original data cube \citep{2016A&A...585A...4C}   to a regular 3D grid, such as typically used by extrapolation methods, and also produced a series of rebinned data cubes with progressively lower resolutions\footnote{All these data cubes with regular spacing as well as standard deviations are available at our project web-site: \url{http://www.ioffe.ru/LEA/SF_AR/files/Magnetic_data_cubes/Bifrost/index.html}}. Referring to Paper I for a more detailed description of the data sets, we give only a very brief overview of the data set here.

%\subsection{Overview of the Data Set}

The highest resolution data cube used in our study was cut out over the height and regridded from the original snapshot and  is  $24~\times~24~\times~12$~Mm, with a grid of $504~\times~504~\times~252$ cells, extending from a nominal photosphere, which corresponds to $Z=0$\ Mm, to 12\ Mm above the photosphere, the equidistant grid spacing of $\simeq$48\ km at all three axes. Other data cubes are the smaller-resolution data with the binning factors $n=2,~3,~4,~6,~7,~9$, which are all multipliers of both 504 and 252; therefore, the lowest resolution, bin\,=\,9, data cube is only $56~\times~56~\times~28$ cells with a spacing of 432~km comparable to the SDO/HMI spatial resolution of 360~km \citep{2012ApJ...748...77S}. The magnetic field at the nominal photosphere is very high, e.g., with the $B_z$ values from $-2,225$ to  $2,081$~G but it is a small-scale field, so that it is only within $\sim\pm60$~G at the chromospheric level at  $Z=2.2$\ Mm.

Following  Paper I, in addition to the nominal photosphere level at the bottom of each data cube, we specify two other important layers. The first of them is a level where the distribution of the plasma parameter $\beta=p_{\rm kin}/p_{\rm B}$, where $p_{\rm kin}$ is the gas pressure and $p_{\rm B}$ is the magnetic pressure, is similar to that in a typical active region (AR) at the sun (we call this layer the $\beta$-photosphere for short), while the other one is a chromospheric level; the heights of all these layers are specified in Table~1 of Paper I. For our testing, we associate the $\beta$-photosphere with the actual AR photosphere, but do not consider extrapolations from the nominal photosphere (see Paper I for details). Although we performed our tests for all binnings, we only present the results for three binning, 3, 6, and 9, which are representative for the entire data set.

\section{Metrics for the testing}
\label{S_dcube_analysis}

In this paper we use a subset of metrics used in Paper I. Some of them \citep{2000ApJ...540.1150W,2006SoPh..235..161S} are needed to evaluate the reconstructed magnetic field force-freeness:
\begin{equation}\label{IM_E01}
    \theta
    =
    \arcsin
    \left(
        \frac{
            \sum_{i}^{N}
            \sigma_{i}
        }
        {
            N
        }
    \right)
    ,
    \quad
    \theta_{j}
    =
    \arcsin
    \left(
        \frac{
            \sum_{i}^{N}
            \left|
                \textit{\textbf{j}}
            \right|_{i}
            \sigma_{i}
        }
        {
            \sum_{i}^{N}
            \left|
                \textit{\textbf{j}}
            \right|_{i}
        }
    \right)
    ,
    \quad
    $$$$
    \sigma_{i}
    =
    \frac{
    \left|
        \textit{\textbf{j}}
        \times
        \textit{\textbf{B}}_{\rm NLFFF}
    \right|_{i}
    }
    {
        \left|
            \textit{\textbf{j}}
        \right|_{i}
        \left|
            \textit{\textbf{B}}_{\rm NLFFF}
        \right|_{i}
    },
\end{equation}
and the solenoidal criterion:
\begin{equation}\label{IM_E02}
    f
    =
    \frac{
        1
    }
    {
        N
    }
    \sum_{i}^{N}
    \frac{
        \left|
            \nabla
            \textit{\textbf{B}}_{\rm NLFFF}
        \right|_{i}
    }
    {
        6
        \left|
            \textit{\textbf{B}}_{\rm NLFFF}
        \right|_{i}
    }
    dx,
\end{equation}
where $\textit{\textbf{B}}_{\rm NLFFF}$ is the reconstructed NLFFF;  $\textbf{\textit{j}}$ is the corresponding electric current density,  the summation is performed over $N$ voxels of the computational subdomain or entire volume (excluding boundaries), \textit{dx} is the grid spacing; $\sigma_i$ is the sine of the angle between the magnetic field and the current density at the $i$-th node of the computational grid; $\theta$ is the angle averaged over all nodes, must be small for a nearly force-free field; $\theta_j$ is a similar metrics but weighted with the electric current that means that contributions form strong currents dominate this metrics.

To assess how close the NLFFF extrapolated data cube is to the corresponding model data cube, we use the ``angular'' metrics similar to Eqn~(\ref{IM_E01}):
\begin{equation}\label{IM_E03}
    \theta_{m}
    =
    \arccos
    \left(
        \frac{
            \sum_{i}^{N}
            \tau_{i}
        }
        {
            N
        }
    \right)
    ,
    \quad
    \theta_{mj}
    =
    \arccos
    \left(
        \frac{
            \sum_{i}^{N}
            \left|
                \textit{\textbf{j}}
            \right|_{i}
            \tau_{i}
        }
        {
            \sum_{i}^{N}
            \left|
                \textit{\textbf{j}}
            \right|_{i}
        }
    \right)
    ,
    \quad
    $$$$
    \tau_{i}
    =
    \frac{
        \textit{\textbf{B}}_{{\rm NLFFF},~i}
        \cdot
        \textit{\textbf{B}}_{i}
    }
    {
        \left|
            \textit{\textbf{B}}_{{\rm NLFFF}}
        \right|_{i}
        \left|
            \textit{\textbf{B}}
        \right|_{i}
    },
\end{equation}
where $\textbf{\textit{B}}$ is the known model field, $\textbf{\textit{j}}$ is the electric current density, computed for reconstructed field, the summation is performed over the voxels of the analyzed volume subdomain, %\textit{dx} is the grid spacing;
$\tau_i$ is the cosine of the angle between the restored and model magnetic field at the $i$-th voxel of the computational grid; $\theta_m$ is the angle averaged over all $N$ voxels of the given subdomain, for a good reconstruction it must be small; $\theta_{mj}$ is a similar metrics but weighted with the restored electric current that ensures that the contribution form voxels with strong electric current dominates this metrics.

For a voxel-to-voxel inspection we compute the local error (residual) $ \Delta_{\alpha}[j]$ and the local relative error $\delta_{\alpha}[j]$ as %, and local normalized residual $\chi_{\alpha}^2[j]$ as %{\gf [Update the equations!!]}

\begin{equation}\label{Eq_NLFFF_err_loc_def}
   \Delta_{\alpha}[j]= B_{\rm NLFFF, \alpha}[j]-\overline{B}_{\alpha}[j], \qquad $$$$
   \delta_{\alpha}[j]= \frac{B_{\rm NLFFF, \alpha}[j]-\overline{B}_{\alpha}[j]}{ \langle\overline{B}_{\alpha}[j]\rangle}, \qquad \alpha=x,~y,~{\rm or}~z
   ,
\end{equation}
where $j$ is the number of a given voxel, $\langle\overline{B}_{\alpha}\rangle=\sqrt{\overline{B}_{\alpha}^2 + \delta B_{\alpha}^2}$, $\overline{B}_{\alpha}[j]$ and $\delta B_{\alpha}^2$ are defined for each binning factor by Equations~(1) and (2) from Paper I.
Here, to compute the relative error, we take into account that after the cube rebinning the magnetic field in each voxel is only known to the accuracy of $\overline{B}_{\alpha}\pm \delta B_{\alpha}$. Thus, in the denominators of $\delta_{\alpha}[j]$ in Eq.~(\ref{Eq_NLFFF_err_loc_def}) we use $\langle\overline{B}_{\alpha}\rangle$ rather than $\overline{B}_{\alpha}$; otherwise, in `singular' points, where $\overline{B}_{\alpha}$ in the denominator is very close to zero, such a metrics would artificially underestimate the accuracy. However, to compute a similar metrics for the absolute value of the magnetic field vector, we do not add any $\delta B$ because the absolute value is never too close to zero in the analyzed volume.

To characterize the extrapolation performance in a given subdomain, which can be, for example, a given layer or the entire data cube, we use the normalized root-mean-square (rms) %\sa{What does \lq\lq rms\rq\rq mean? We need to notate it.}
residual $\Delta_{\rm rms}$ %and the normalized rms error $\delta_{\rm rms}$ %, and 'effective $\chi^2$' metrics
defined as

\begin{equation}\label{Eq_NLFFF_NormRes_def}
   \Delta_{\rm rms, \alpha}=\sqrt{\frac{\sum\limits_{j=1}^{N_{\rm vox}}  \Delta_{\alpha}^2[j]}{\sum\limits_{j=1}^{N_{\rm vox}} \overline{B}_{\alpha}^2[j]}}
   %\frac{(B_{\rm NLFFF, \alpha}[j]-\overline{B}_{\alpha}[j])^2}{\delta B_{\alpha}^2[j]}
   , \qquad \alpha=x,~y,~{\rm or}~z
   ,
\end{equation}
where the summation is performed over the subdomain of the data cube\footnote{The extrapolated data cubes are the subject of data sharing by request to the authors.} %\footnote{\gf All reconstructed data cubes obtained in our study are available at our project web-site: \url{http://www.ioffe.ru/LEA/SF_AR/files/Magnetic_data_cubes/Extrapolations-Bifrost/index.html}.} {\gf [GF: will we put the cubes on-line?]}
used for the analysis; $N_{\rm vox}$ is the total number of voxels in the selected subdomain. %For example, a few boundary layers can be discarded while metrics are computed to reduce the effect of the boundaries, where the extrapolated field is expected to stronger deviate from the true field.
The normalized rms residual, Eq.~(\ref{Eq_NLFFF_NormRes_def}), gives more weight to the voxels, where the magnetic field is strong. %In contrast, metric (\ref{Eq_NLFFF_err_def}) gives equal weight to any voxel, with either strong or weak field. %The $\chi^2$ metrics weights the voxel in accordance with the uncertainty to which the magnetic field is known in the given voxel.

In addition, we quantify the reconstructed magnetic connectivity by computing the magnetic field lines in the reconstructed and original data cubes and computing the largest deviations between the corresponding field lines. Finally, to assess the picture of the reconstructed electric currents we compute LOS-integrated maps of the electric current for visual inspection.

\section{NLFFF Reconstruction with Chromospheric Magnetic Field Data}
\label{S_nlfff}

Although there exists a number of NLFFF reconstruction methods, see, e.g., brief overview in \citet[][Section 5.3.3]{2005psci.book.....A} and \citet{2015ApJ...811..107D}, various versions of the optimization method  \citep{2000ApJ...540.1150W} are, perhaps, the most easily adjustable to adding different constraints in the modeling volume such as chromospheric magnetic field data, which can be incomplete, pertain to different chromospheric heights etc. Accordingly, in this study we employ corresponding modifications of the two reconstruction codes tested earlier in Paper I: AS code, following \citet[][]{2004SoPh..219...87W} and IM code, following \citet[][]{2009SoPh..257..287R}.

\subsection{Modifications to the AS code}
\label{S_modif_AS}

In both implementations used in this paper, the NLFFF reconstructions are performed following the optimization method \citep{2000ApJ...540.1150W}. The main idea of the optimization method is to transform some trial configuration of the magnetic field (usually a potential extrapolation from the bottom boundary) to a final force-free field configuration. This is achieved by minimization of the following, positively defined, functional:

\begin{equation}\label{Eq_nlfff_func_chr}
L=\int\limits_{V} \left[B^{-2}\left[ [\nabla\times\mathbf{B}]\times\mathbf{B}\right]^2 \right.
       \left.
+ |\nabla\cdot\mathbf{B}|^2\right] w(x,y,z) dV
+  $$$$  \nu\int\limits_{\widetilde{V}}(\mathbf{B}-\mathbf{B}_{\rm obs})\cdot\mathbf{W}\cdot(\mathbf{B}-\mathbf{B}_{\rm obs}) \frac{dV}{\Delta x\Delta y}
,
\end{equation}
%{\gf [Q: in fact, we can add an arbitrary volume subdomain. How we take care about dimensions of these two terms -- volume and surface integrals?]}
where $w(x,y,z)$ is a 'weight' function, the same as in Paper I, $\mathbf{B}_{\rm obs}$ describes the additional (chromospheric and / or coronal) constraints available in a volume subdomain $\widetilde{V}$, $\mathbf{W}$ describes weights of each vector component, $\nu$ is the user-defined Lagrangian multiplier; in this study we adopt $\nu=1$ in all cases. % {\gf [How the multiplier is defined?]}.
According to a variety of possible constraints at the chromosphere and / or corona, we implemented, as independent options, the vector, the LOS, and the absolute value of the magnetic field, which can exist simultaneously either in the same or different volume subdomains; weights $0\leq{W}\leq1$  characterize the uncertainty of the given constraint and, thus, how strongly it affects the solution. {\gf In particular, these uncertainties can include the uncertainty of the  chromospheric/coronal height, to which a given constraint is formally assigned.} Although Eqn~(\ref{Eq_nlfff_func_chr}) has a form similar to that in \citet{2010A&A...516A.107W}, there is essential difference between those two cases. Indeed, in \citet{2010A&A...516A.107W} the second term represents a surface integral over the photospheric boundary and so allows the solution to deviate from the bottom boundary condition, which is only known within a certain accuracy. In our approach, we assume that the photospheric magnetic field is known with a much higher accuracy than any additional magnetic constraints; so the bottom boundary of the reconstructed data cube perfectly matches the boundary condition (no deviation is allowed), while the second integral pertains to an arbitrary subdomain in the volume, for example, a chromospheric layer, where the additional magnetic field data are available.

%--Give extended equation.
%
%--Describe, how the conditions  $>$ or $<$ are implemented.
%
%--How the oscillatory behavior was overcome.
%
%--Other technical details.

After numerous tests with this implementation, we found that adding constraints at the chromospheric level has a positive impact on the solution just above the chromosphere but this positive effect vanishes soon at the higher heights. This happens due to two reasons. Firstly, the functional is effectively weighted by the bottom part of the data cube, where both the magnetic field and electric current are strong. And secondly, the top and side boundaries are fixed from the potential field extrapolation based on the initial photospheric boundary. To get rid of these two problems, we perform the reconstruction in three steps: (i) apply the optimization code with all available additional constraints to a subdomain with a height needed to inscribe all these constraints, then (ii) cut a plane ``chromospheric'' layer and perform the standard extrapolation starting from this almost force-free boundary, and finally (iii) glue the bottom part from step (i) to the data cube such as the final data cube covers all the heights starting from the bottom photospheric vector boundary. We note that this approach is also helpful when the additional constraints pertain to various chromospheric and coronal heights, rather than to a plane surface.

\subsection{Modifications to the IM code}

%Modifications to the IM code were performed differently.
%
%--Describe, how exactly this was done and how the uncertainties are taken into account.

In contrast to the AS implementation, the IM implementation of the optimization method does not include any modification of the main functional. The functional has its original form as it was introduced in \citet{2000ApJ...540.1150W}:

\begin{equation}\label{IM_E04}
	L
	=
	\int \limits_{V}
		\left[
			B^{-2}
			\left[
				\left[
					\nabla
					\times
					\mathbf{B}
				\right]
				\times
				\mathbf{B}
			\right]^2
			+
			|
				\nabla
				\cdot
				\mathbf{B}
			|^2
		\right]
		dV.
\end{equation}
Likewise, all equations for evaluating magnetic field in the volume and on the lateral and top boundaries of the computational domain remain unchanged, while the additional constraints are straightforwardly applied at each iteration. In the present study, we do not consider any uncertainty of the additional constraints. % noise influence, so additional inner constraints can be taken into account in a quite simple way.
Thus, any available additional information on magnetic field must be preserved during the NLFFF reconstruction procedure.
%\textbf{\sout{To ensure that, at every consequent iteration, when the magnetic field is recalculated everywhere according to evolutionary equations, it is adjusted at those voxels where some additional information is available.}}
{To ensure that the extrapolated magnetic field satisfies additional inner constraints, we adjust its components at every consequent iteration   at those voxels where some additional information is available after recalculating magnetic field everywhere according to evolutionary equations.}

In particular, when information on the magnetic field component along a specified line-of-sight direction is provided, the newly recalculated field vector is decomposed into line-of-sight and tangential components. Then, the former is set to be equal to the provided constraints, while the latter remains unchanged. In the case when the information on the vector magnitude is provided, the newly recalculated field vector is renormalized according to the constraints while preserving its direction. When both types of inner constraints are provided then first the line-of-sight constraints are applied, and after this, the tangential component is changed to match the magnitude constraints.

\section{NLFFF Reconstruction Tests}

\begin{figure*} %[!t]
\centering
\includegraphics[width=0.58\textwidth,clip]{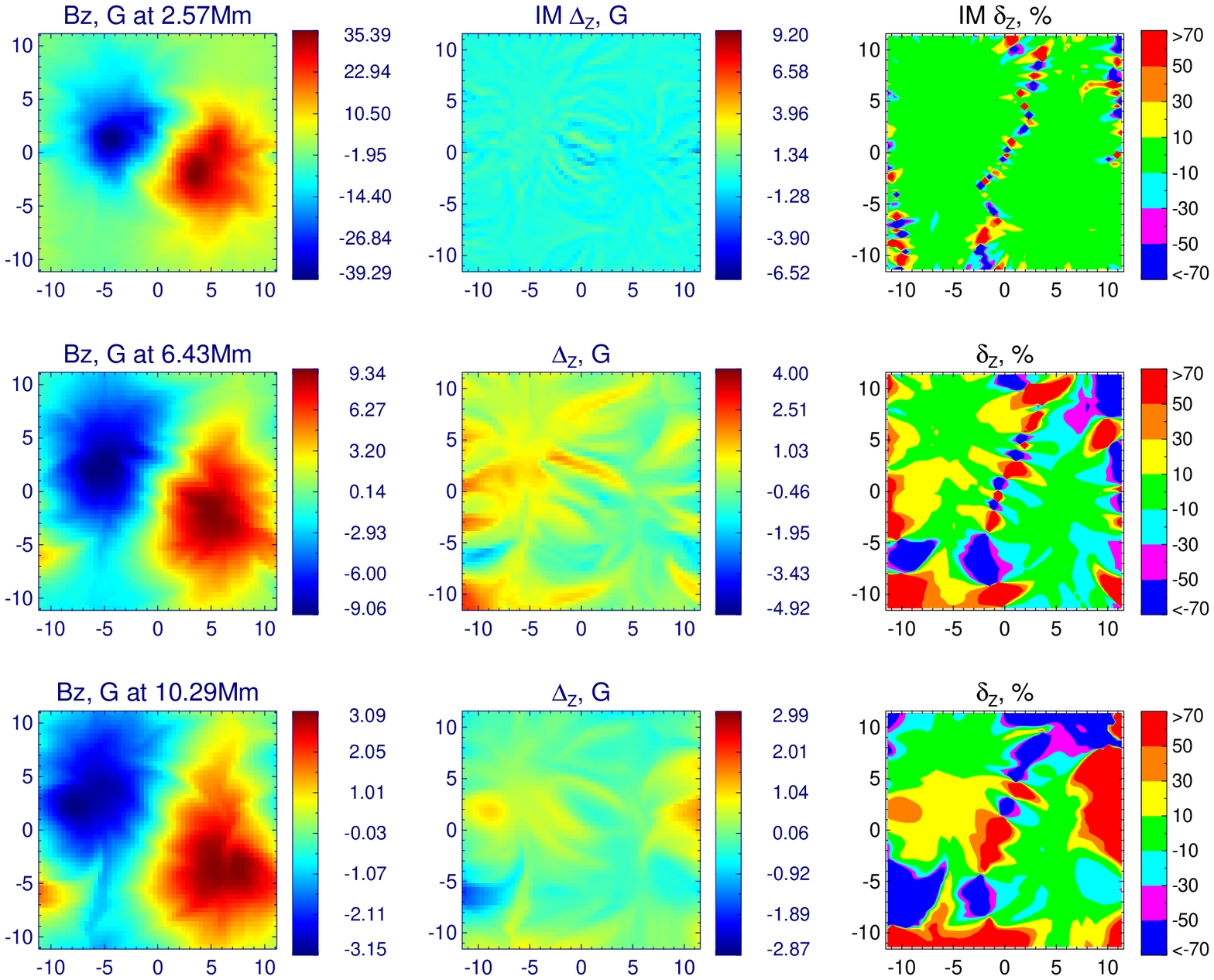}\quad
\includegraphics[width=0.375\textwidth,clip]{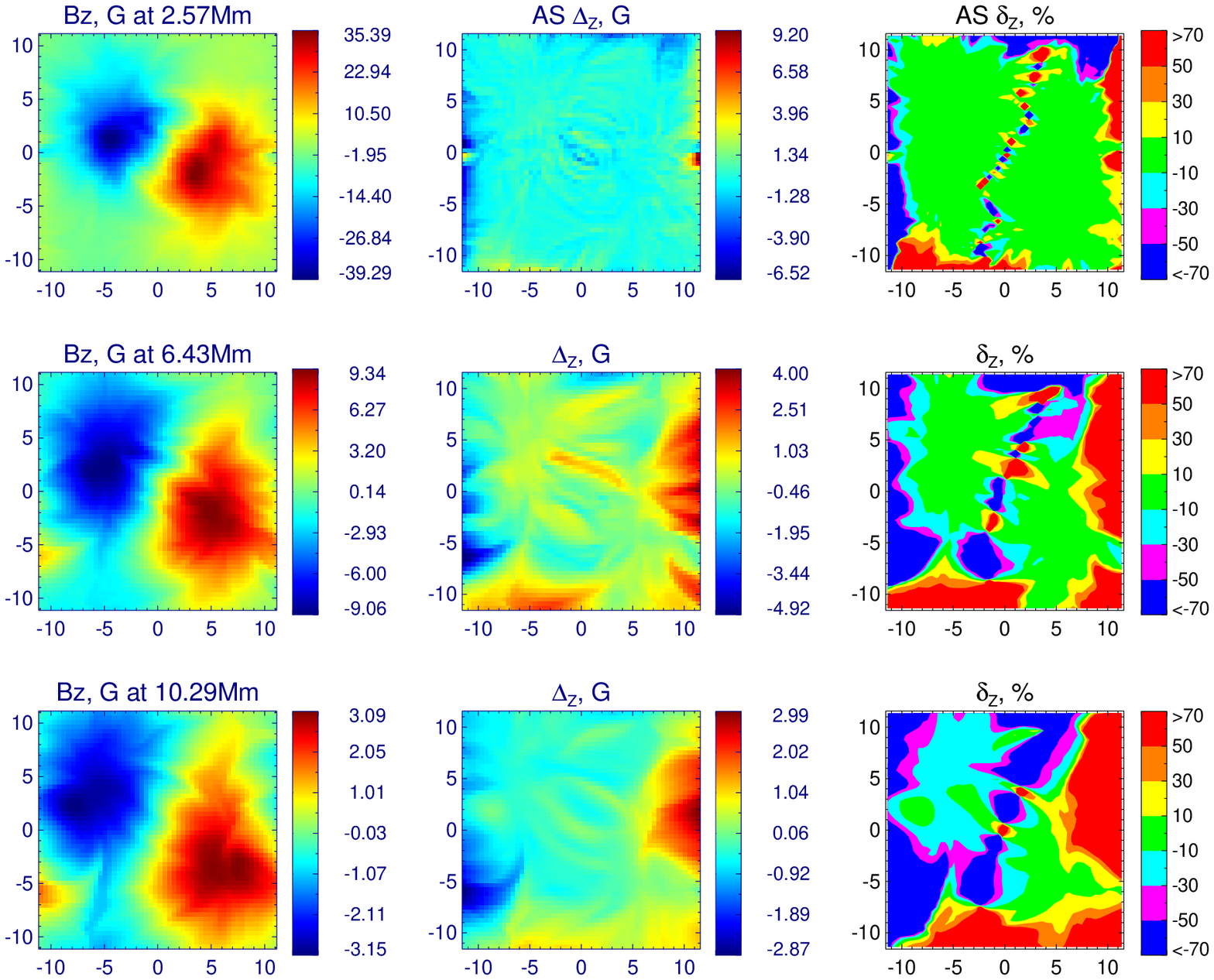} %,bb=230 17 581 453,
\caption{\label{f_bifrost_385_bin9_chrNLFFF_Bz} The model $B_z$ field distributions (left column) and performance of the IM (next two columns) and AS (two right columns) NLFFF extrapolations ($B_z$ component) from the photospheric level with inclusion of full vector information at the chromospheric level (bin\,=\,9) at three levels (their height are shown at the panel titles); {second and fourth columns: residual between the extrapolated and the model field; third and fifth column: relative error.  The relative error is bigger along the `neutral lines,' where the field is close to zero. The results for other components and other binning factors are similar to those shown in this figure. The animated version of this figure shows the same information but for all layers of the reconstructed $B_z$ data cubes separately for the IM and AS extrapolations. Each frame of the animations shows four panels at a given layer: (a) the model field, (b) the restored field, (c) the residual, and (d) the relative error.}
%(5693.874 s).
 }
\end{figure*}

We expect  the extrapolations  to improve  due to adding constraints at the chromospheric level, where the magnetic field is close to the force-free state. {\gf Here we aim to evaluate the potential \textit{algorithmic} improvement of the NLFFF reconstruction due to adding chromospheric constraints. Thus, we do not explicitly consider the height uncertainties of the chromospheric constraints.}
The index of the voxel layer where the chromospheric constraints are applied  depends on the model resolution (bin-factor) and is tabulated in Table~1 of Paper I.
To assess the potential improvement of NLFFF reconstructions due to the additional chromospheric constraints, we produced a series of the NLFFF extrapolated datacubes by both AS and IM codes, for various bin-factors as in Paper I, where various kinds or combinations of the chromospheric constraints have been employed in addition to the vector ($\beta-$)photospheric boundary condition. According to the types of anticipated chromospheric constraints in the real data, we consider the cases when either the LOS component, or the absolute value, or the full vector of the magnetic field is available. We also consider an option when a combination of the LOS and absolute value of the magnetic field are known. For the cases when the LOS component is involved, we considered both the 'top-view' geometry and also two cases of an oblique viewing angles. In what follows, we present results of global and layer-by-layer assessment of the quality of reconstructions,  the accuracy of connectivity reconstruction, and the fidelity of the electric current restoration.

\subsection{Overview and the Global Assessment}

Qualitatively, the overall behavior of the reconstructed magnetic field is similar to that revealed in Paper I,  although quantitative metrics improve in most cases when additional constraints are used.
Figure~\ref{f_bifrost_385_bin9_chrNLFFF_Bz} shows local error (residual) and normalized local error for the case when the magnetic vector data are available at both photospheric and chromospheric layers. The metrics are comparable for both IM and AS codes, although the AS metrics are marginally better in the inner part of the data cube. The animated version of the figure shows these metrics at all layers of the data cube. Interesting, that the reconstruction yields an appropriate accuracy even at the lowest layers of the data cube, where the field deviates from the force-free state.

To globally assess the  {\gf force-freeness} of the reconstructions, we employ the angular metrics described by Eqns.~(\ref{IM_E01}) and (\ref{IM_E03}), which are  summarized in Table~\ref{IM_T01} for the top view and in Table~\ref{IM_T02} for two side views. It is interesting to note that our two codes behave differently for different sets of the additional constraints at the chromospheric layer. For example, when only $B_z$ component is available at the chromospheric layer, the AS code clearly outperforms the IM one, while both codes show a comparable performance, when $|\textbf{B}|$ is available. In contrast, when two or three components of the magnetic field are available at the chromospheric layer, the IM code gives better results than the AS code. The reason for that is that the IM code employs equations for the side and top boundaries. Thus, when at least two chromospheric components are available, then a correct force-free information propagates by these equations to the higher heights, which helps to improve the reconstruction. It turns out that having only one chromospheric component is insufficient to do that; thus, the weighted optimization approach with the buffer zone comprising 10\% layers at the top and side boundaries in the AS code \citep[see][for the details]{2004SoPh..219...87W, 2017ApJ...839...30F} is more appropriate in those cases.
This tendency holds also for the oblique viewing angle,  Table~\ref{IM_T02}. {\gf We also checked the solenoidal condition: in all cases it holds, within a factor of 2, at the same level as in the original model reported in Table 1 in Paper I; see Table~\ref{T_divless} for details. It is interesting that for larger bin-factors, the solenoidal condition holds better for the reconstructed cubes compared with the original binned model. The reason is that the binned model is obtained from the original one by a direct averaging without taking care of the solenoidal condition, while the extrapolations look for the best solution consistent with (i) bottom boundary condition, (ii) force-free condition, and (iii) solenoidal condition.  }

\begin{deluxetable*}{ c c | c c c c | c c c c | c c c c | c c c c }
	
	\tabletypesize{ \small }
	\tablewidth{ 0pt }
	\tablecaption{ Performance of the magnetic field reconstruction methods with different types of inner constraints \label{IM_T01} }
	\tablehead{
		\colhead{ Bin } & \colhead{ Impl } & \multicolumn{4}{c}{ $B _{z}$ } & \multicolumn{4}{c}{ $\left| \mathbf{B} \right|$ } & \multicolumn{4}{c}{ $B _{z}$ \&  $\left| \mathbf{B} \right|$ } & \multicolumn{4}{c}{ $\mathbf{B}$ } \\
		& & \colhead{ $\theta^{\circ}$ } & \colhead{ $\theta _{j}^{\circ}$ } & \colhead{ $\theta _{m}^{\circ}$ } & \colhead{ $\theta _{mj}^{\circ}$ } & \colhead{ $\theta^{\circ}$ } & \colhead{ $\theta _{j}^{\circ}$ } & \colhead{ $\theta _{m}^{\circ}$ } & \colhead{ $\theta _{mj}^{\circ}$ } & \colhead{ $\theta^{\circ}$ } & \colhead{ $\theta _{j}^{\circ}$ } & \colhead{ $\theta _{m}^{\circ}$ } & \colhead{ $\theta _{mj}^{\circ}$ } & \colhead{ $\theta^{\circ}$ } & \colhead{ $\theta _{j}^{\circ}$ } & \colhead{ $\theta _{m}^{\circ}$ } & \colhead{ $\theta _{mj}^{\circ}$ }
	}
	
	\startdata
	%    	\\
	\multirow{2}{*}{ 3 } & IM & 20.0 & 14.9 & 29.0 & 16.6 & 15.6 & 13.8 & 24.7 & 15.8 & 13.7 & 14.1 & 15.8 & 10.9 & 12.8 & 13.6 & 15.2 & 10.1 \\
	& AS & 28.4 & 17.5 & 25.7 & 14.1 & 25.3 & 17.4 & 26.2 & 15.3 & 25.8 & 17.3 & 24.5 & 13.3 & 26.5 & 17.0 & 23.9 & 12.2 \\
	%& IM/AS & 21.7 & 14.9 & 27.1 & 15.9 & 16.6 & 14.3 & 24.2 & 17.2 & 15.6 & 14.1 & 20.0 & 14.0 & - & - & - & - \\
	\\
	%\multirow{3}{*}{ 4 } & IM & 22.1 & 17.8 & 35.8 & 20.8 & 16.1 & 16.4 & 25.8 & 16.5 & 14.5 & 16.9 & 15.6 & 12.0 & 13.7 & 16.3 & 14.9 & 11.0 \\
%	& AS & 29.2 & 19.1 & 25.6 & 14.5 & 25.7 & 18.4 & 25.9 & 15.5 & 26.6 & 19.2 & 24.4 & 14.2 & 27.1 & 18.2 & 23.8 & 13.6 \\
%	& tmp & 22.8 & 16.4 & 27.6 & 16.9 & 17.5 & 15.3 & 24.5 & 18.0 & 17.3 & 15.9 & 20.5 & 15.7 & - & - & - & - \\
%	\\
	\multirow{2}{*}{ 6 } & IM & 20.5 & 16.5 & 34.8 & 19.2 & 18.8 & 16.5 & 22.8 & 13.4 & 16.1 & 15.7 & 15.7 &  9.2 & 14.9 & 14.9 & 14.5 &  8.0 \\
	& AS & 25.2 & 15.4 & 24.4 & 12.2 & 23.2 & 15.0 & 24.9 & 14.1 & 23.9 & 15.5 & 23.8 & 12.3 & 26.3 & 16.3 & 23.2 & 10.9 \\
	%& IM/AS & 20.1 & 13.1 & 22.6 & 13.5 & 17.3 & 12.9 & 21.0 & 14.6 & 16.6 & 12.7 & 18.9 & 12.5 & - & - & - & - \\
	\\
	%\multirow{3}{*}{ 7 } & IM & 19.7 & 15.0 & 22.5 & 11.6 & 17.9 & 14.4 & 18.3 & 10.3 & 15.9 & 13.2 & 14.1 &  7.5 & 14.6 & 12.8 & 13.1 &  6.5 \\
%	& AS & 23.4 & 12.7 & 23.5 & 11.1 & 23.0 & 12.6 & 23.8 & 12.3 & 23.5 & 13.0 & 23.0 & 11.3 & 27.0 & 14.1 & 22.3 & 9.8 \\
%	& tmp & 17.0 & 10.7 & 20.1 & 11.2 & 16.0 & 10.6 & 19.1 & 12.4 & 15.5 & 10.4 & 17.8 & 10.3 & - & - & - & - \\
%	\\
	\multirow{2}{*}{ 9 } & IM & 23.9 & 19.2 & 25.2 & 14.2 & 20.0 & 18.3 & 19.7 & 10.9 & 17.6 & 16.9 & 13.5 &  7.6 & 16.1 & 16.1 & 12.9 &  6.8 \\
	& AS & 26.2 & 16.6 & 24.1 & 11.4 & 25.1 & 16.2 & 24.4 & 12.8 & 25.5 & 16.7 & 23.5 & 11.8 & 30.0 & 18.0 & 22.8 & 10.3 \\
	%& IM/AS & 21.1 & 14.6 & 21.5 & 12.7 & 19.2 & 14.4 & 20.0 & 14.0 & 18.6 & 14.2 & 18.0 & 12.0 & - & - & - & - \\
	%        \\
	\enddata
	
	\tablecomments{ Bin\,--\,is binning factor. Impl\,--\,is implementation of the optimization method. Four subsequent large columns contain numerical characteristics of reconstructed field with use of corresponding type of inner constraints on the chromosphere level. %IM/AS means the reconstruction performed by the IM code starting from the chromospheric condition taken from the AS reconstruction.
}
	
\end{deluxetable*}

\begin{figure*}
	\centering
	\includegraphics[width=1.0\textwidth]{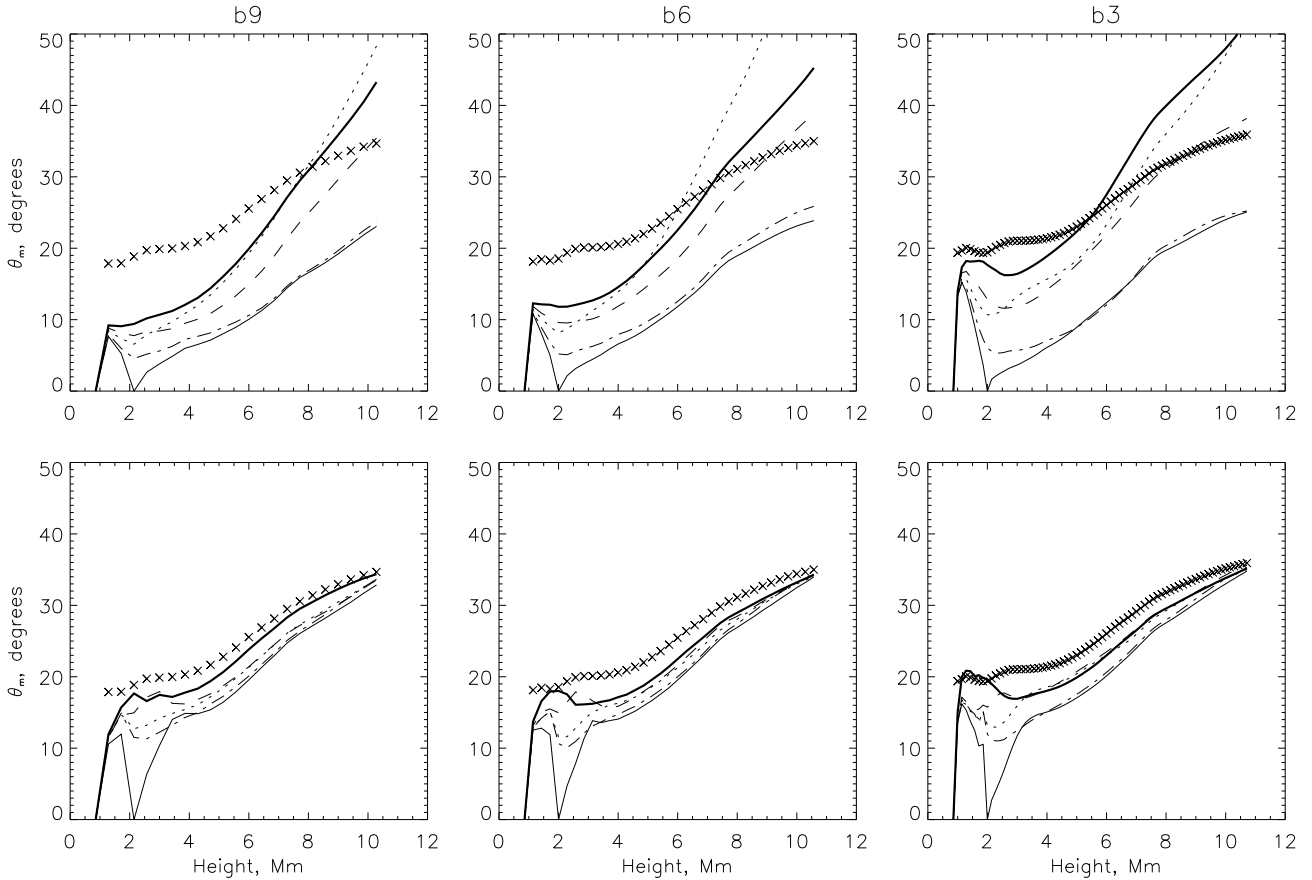}\\
	\caption{\label{theta_m_layers_u}
		%-= AS functional =-
		Distribution of $\theta _{m}$ parameter, computed for horizontal layers at given heights with use of different inner constraints on the chromosphere for three  binnings\,=\,3,\,6,\,9. The first row shows the results obtained with the IM code. The second row---with the AS code.  "x" symbols show the metrics fr the initial potential extrapolation. Thick solid line: no inner constraints (reference case). Dotted line: $B _{z}$. Dashed line: $\left| \mathbf{B} \right|$. Dash-dotted line: $B _{z}$ \& $\left| \mathbf{B} \right|$. Thin solid line: $\mathbf{B}$.
	}
\end{figure*}

\begin{deluxetable*}{ c c | c c c c | c c c c | c c c c | c c c c }
	
	\tabletypesize{ \small }
	\tablewidth{ 0pt }
	\tablecaption{ Performance of the magnetic field reconstruction methods using magnetic field component along predetermined line-of-sight direction as inner constraint \label{IM_T02} }
	\tablehead{
		\colhead{ Bin } & \colhead{ Impl } & \multicolumn{4}{c}{ $B _{W00N30}$ } & \multicolumn{4}{c}{ $B _{W00N30}$ \&  $\left| \mathbf{B} \right|$ } & \multicolumn{4}{c}{ $B _{W30N30}$ } & \multicolumn{4}{c}{ $B _{W30N30}$ \&  $\left| \mathbf{B} \right|$ } \\
		& & \colhead{ $\theta^{\circ}$ } & \colhead{ $\theta _{j}^{\circ}$ } & \colhead{ $\theta _{m}^{\circ}$ } & \colhead{ $\theta _{mj}^{\circ}$ } & \colhead{ $\theta^{\circ}$ } & \colhead{ $\theta _{j}^{\circ}$ } & \colhead{ $\theta _{m}^{\circ}$ } & \colhead{ $\theta _{mj}^{\circ}$ } & \colhead{ $\theta^{\circ}$ } & \colhead{ $\theta _{j}^{\circ}$ } & \colhead{ $\theta _{m}^{\circ}$ } & \colhead{ $\theta _{mj}^{\circ}$ } & \colhead{ $\theta^{\circ}$ } & \colhead{ $\theta _{j}^{\circ}$ } & \colhead{ $\theta _{m}^{\circ}$ } & \colhead{ $\theta _{mj}^{\circ}$ }
	}
	
	\startdata
	%    	\\
	\multirow{2}{*}{ 3 } & IM & 19.7 & 14.5 & 27.5 & 16.4 & 13.9 & 13.8 & 18.6 & 12.6 & 19.0 & 14.2 & 27.0 & 15.8 & 14.9 & 14.2 & 17.6 & 11.8 \\
	& AS & 25.8 & 13.8 & 25.9 & 16.6 & 23.2 & 13.7 & 25.0 & 16.6 & 27.0 & 13.9 & 25.7 & 15.9 & 23.4 & 13.6 & 25.1 & 16.2 \\
	\\
	\multirow{2}{*}{ 6 } & IM & 19.3 & 16.5 & 32.1 & 18.3 & 16.1 & 15.6 & 17.7 &  9.9 & 19.6 & 16.2 & 30.3 & 16.0 & 17.7 & 16.3 & 18.0 &  9.7 \\
	& AS & 24.2 & 14.6 & 24.4 & 12.5 & 23.4 & 14.7 & 23.9 & 12.7 & 24.8 & 14.7 & 24.4 & 12.6 & 23.7 & 14.7 & 23.9 & 12.6 \\
	\\
	\multirow{2}{*}{ 9 } & IM & 23.0 & 19.0 & 22.8 & 12.2 & 17.6 & 17.0 & 15.1 &  8.3 & 22.2 & 18.4 & 24.7 & 14.0 & 19.4 & 17.8 & 15.1 &  8.2 \\
	& AS & 25.2 & 16.2 & 24.3 & 12.0 & 25.2 & 16.6 & 23.7 & 12.0 & 25.7 & 16.2 & 24.1 & 11.7 & 25.9 & 16.7 & 23.6 & 11.7 \\
	%        \\
	\enddata
	
	\tablecomments{ Bin\,--\,is binning factor. Impl\,--\,is implementation of the optimization method. Four subsequent large columns contain numerical characteristics of reconstructed field with use of corresponding type of inner constraints on the chromosphere level. }
	
\end{deluxetable*}

\subsection{Layer-by-layer Assessment}

Figure~\ref{theta_m_layers_u} shows how one of those angular metrics, the $\theta_m$ parameter computed for a given layer, changes with height for various cases.
In the case of the IM code, the results are notably different for various combinations of the additional constraints. Adding just one $B_z$ chromospheric component does not help to improve the NLFFF reconstruction too much: this metrics is overall similar to that without any chromospheric data. Even though it marginally improves the reconstruction at low heights, the metrics is getting worse at the higher heights. Both NLFFF reconstructions (without any additional constraint and with $B_z$ only) are less accurate at the higher heights than the potential extrapolation. In contrast, adding the absolute value $|\textbf{B}|$ improves this metrics noticeably; in fact, making it comparable with that for the AS code for the  $|\textbf{B}|$ case. Having the full vector at the chromospheric layer further improves the metrics noticeably. Remarkable is that having just two vector components, the combination of  $|\textbf{B}|$ and $B_z$ has almost the same positive effect on the reconstruction as having the full vector. This happens because the vector components are linked to each other by the equation $\nabla\cdot\textbf{B} =0$; thus, having two component is already sufficient to significantly constrain the third one.

In the case of the AS code the overall improvement follows the same trend but it is less pronounced quantitatively. The reconstruction quality is always better than in the case of the potential extrapolation; adding even one component improves the reconstruction noticeably, while adding more components results in less prominent improvement compared with the IM code. This is the outcome of the adopted construction of the solution, when the side and top boundary conditions are fixed from the potential extrapolation and the buffer zone is employed at the boundary domains as described in Section~\ref{S_modif_AS}.

\begin{figure}
%  \qquad  \quad \textbf{(a) IM:  $B_z$ }\qquad \qquad\ \ \ \textbf{(b) AS:  $B_z$} \qquad \qquad\quad  \ \ \textbf{(c) IM:   $|\mathbf{B}|$} \qquad \qquad \quad \ \  \textbf{(d) AS:  $|\mathbf{B}|$} \\
\centering
\includegraphics[width=0.98\columnwidth]{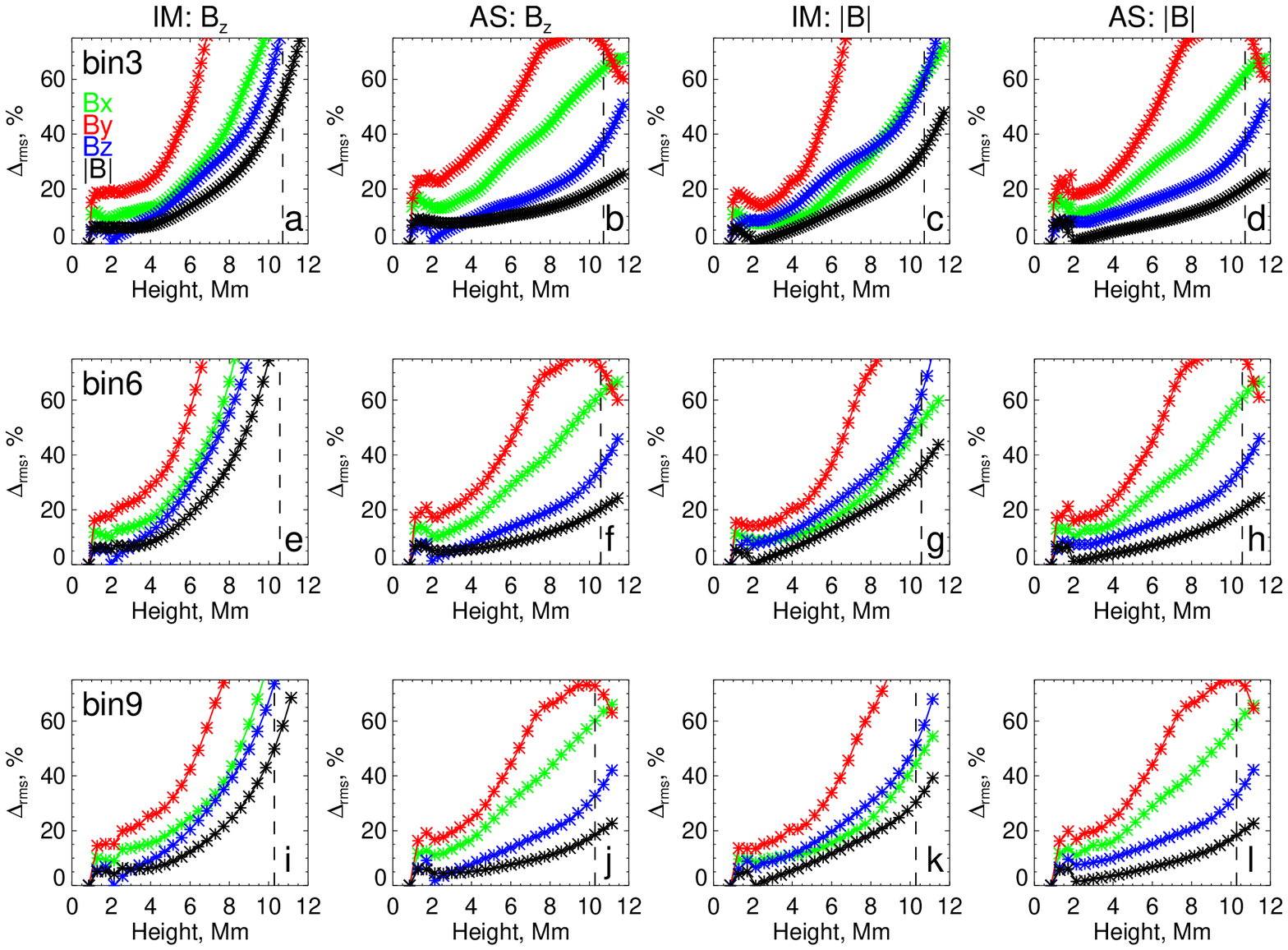}
\caption{\label{f_bifrost_385_NLFFF_deltaVSheight_BBz}
%(5693.874 s).
Relative rms residual in a layer as a function of height for the NLFFF reconstructions obtained for three different resolutions, bin\,=\,3 (top row), bin\,=\,6 (middle row), and bin\,=\,9 (bottom row), using two methods, IM \& AS, from the  photospheric level with inclusion of either the $B_z$ component (the LOS component for top view) or the absolute value $|\mathbf{B}|$ of the magnetic field at the chromospheric level. The side buffer zones are discarded everywhere, while the height of the top buffer zone is shown by the dashed vertical line.
 }
\end{figure}

\begin{figure}
  %\quad\  \textbf{(a) IM:  $B_z$ \& $|\mathbf{B}|$ }\quad \quad\ \  \textbf{(b) AS:  $B_z$ \& $|\mathbf{B}|$} \qquad \quad\ \ \ \ \textbf{(c) IM:   $\mathbf{B}$} \qquad \qquad \quad \ \  \textbf{(d) AS:  $\mathbf{B}$} \\
\centering
\includegraphics[width=0.98\columnwidth]{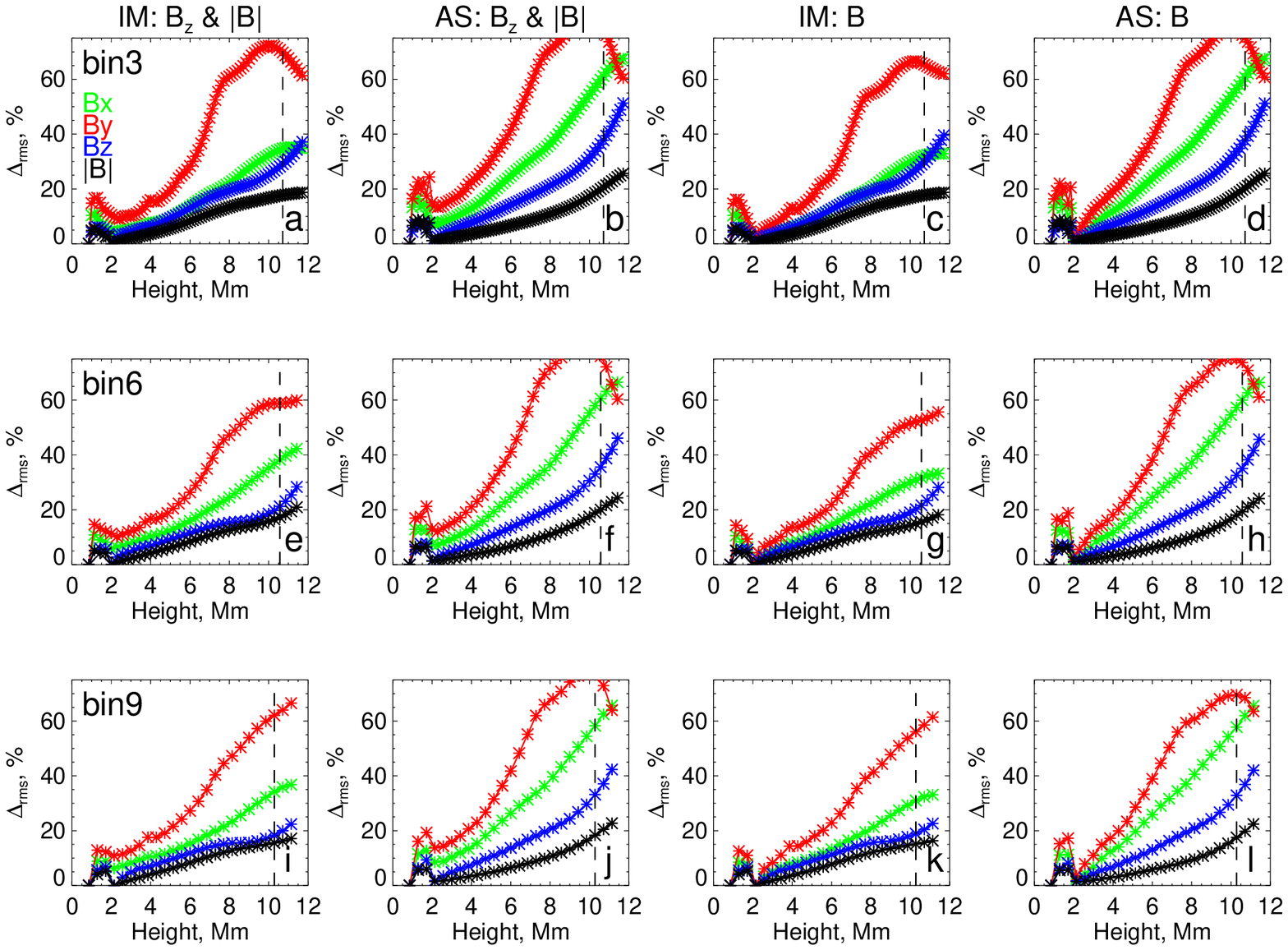}
\caption{\label{f_bifrost_385_NLFFF_deltaVSheight_B_vecB}
%(5693.874 s).
Relative rms residual in a layer as a function of height for the NLFFF reconstructions obtained for three different resolutions, bin\,=\,3 (top row), bin\,=\,6 (middle row), and bin\,=\,9 (bottom row), using two methods, IM \& AS, from the  photospheric level with inclusion of either a combination of the $B_z$ component {and} the absolute value $|\mathbf{B}|$ {or} the full vector $\mathbf{B}$ information at the chromospheric level. The side buffer zones are discarded everywhere, while the height of the top buffer zone is shown by the dashed vertical line.
 }
\end{figure}

The angular metrics considered above do not tell us how well the various components of the magnetic filed are reproduced. A straightforward way of testing if the reconstructed values are correlated with the model ones, would be to produce the corresponding scatter plots and compare the so obtained 2D cloud of the data point with the ideal correlation $y=x$. This was checked and confirmed (Figures~9\,--\,11 in Paper I) for reconstructions started from both photospheric and chromospheric boundaries, and this is also the case for the present study; thus, we do not show those scatter plots.  To further quantify the reconstruction accuracy (scatter around the $y=x$ line) of the field components (as well as the absolute value), we employ the residual metrics, described by Eqn.~(\ref{Eq_NLFFF_NormRes_def}). The results obtained for a subset of our modeling data cubes, specifically, for bin\,=\,3,\,6,\,and 9, are shown in Figures~\ref{f_bifrost_385_NLFFF_deltaVSheight_BBz} and \ref{f_bifrost_385_NLFFF_deltaVSheight_B_vecB}.

Figure~\ref{f_bifrost_385_NLFFF_deltaVSheight_BBz} summarizes the metrics for the case when only one chromospheric component is available, either $B_z$ or $|\textbf{B}|$. The errors increase rapidly with the height for the IM code if only $B_z$ component is available (the first column), while the situation improves if the absolute value $|\textbf{B}|$ is available (the third column). The AS code performs better than the IM one in this plot showing only a modest improvement between the $B_z$ (second column) and $|\textbf{B}|$ (fourth column) cases. In particular, the quality of the $|\textbf{B}|$ reconstruction improves if the absolute value $|\textbf{B}|$ is available at the chromospheric level. $B_z$ component is typically reconstructed less accurately than the $|\textbf{B}|$ one, $B_x$ is further less accurately, and the $B_y$ one is the least accurately.

Figure~\ref{f_bifrost_385_NLFFF_deltaVSheight_B_vecB} summarizes the metrics for the case when more than one chromospheric component is available, either $B_z$ and $|\textbf{B}|$, or the full vector. Here, the quality of the IM reconstruction is greatly improved; it  works better than the AS code for most components. The quality of the $|\textbf{B}|$ reconstruction is, however, comparable for both codes. At a given height, either of the codes can work better than the other; the metrics are tabulated for the quantitative analysis in Tables~\ref{table_rms_res}\,--\,\ref{table_rms_res_4}.

\begin{figure}
\includegraphics[width=0.98\columnwidth]{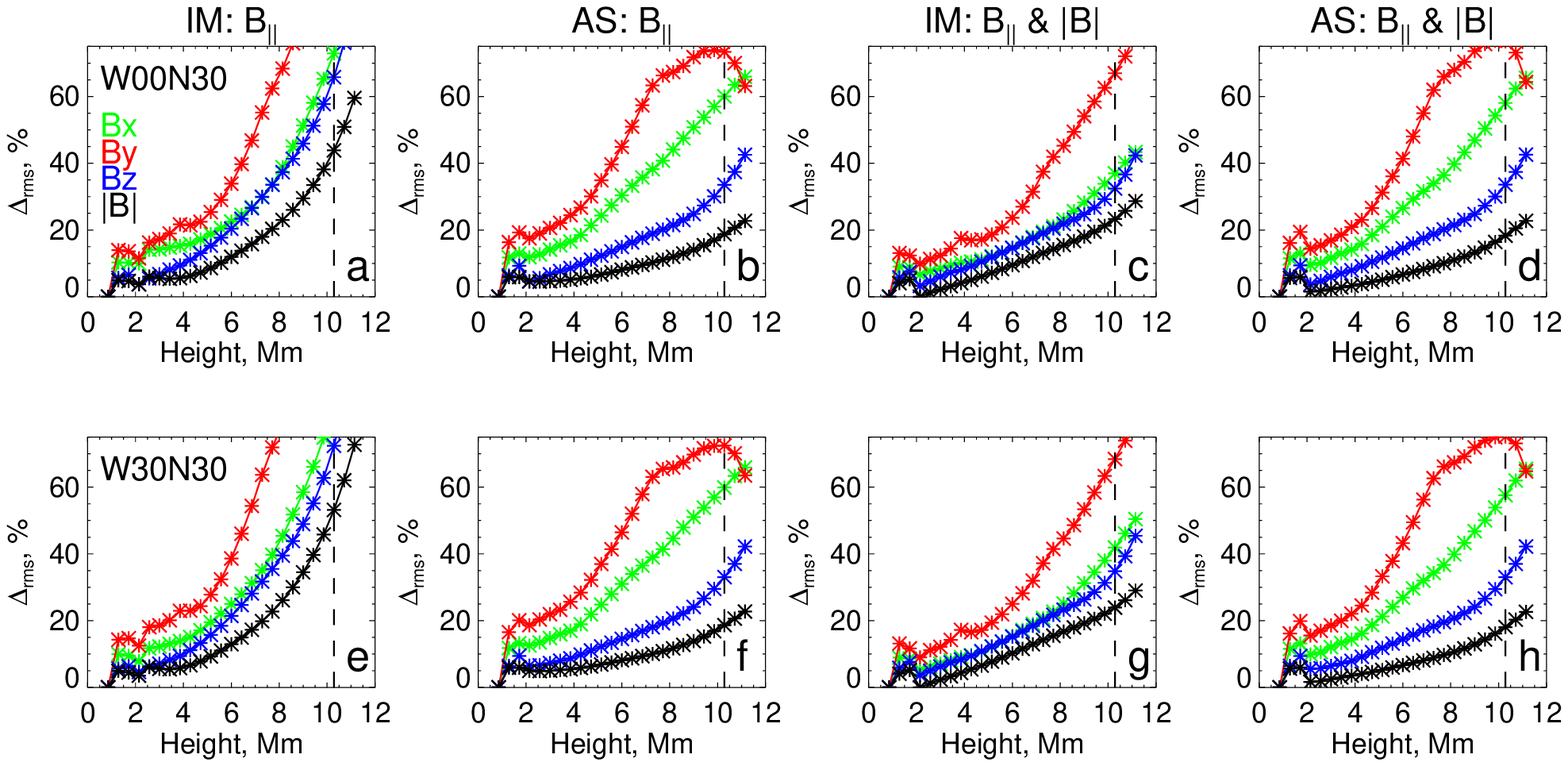}
\caption{\label{f_bifrost_385_NLFFF_deltaVSheight_B_vecB_offCenter}
%(5693.874 s).
Relative rms residual in a layer as a function of height for the bin\,=\,9 NLFFF reconstructions obtained using two methods, IM \& AS, from the  photospheric level with inclusion of either the $B_{||}$ component (the LOS component) or the combination of the LOS component and the absolute value $|\mathbf{B}|$ of the magnetic field at the chromospheric level for two different locations of the data cube at the solar surface: 30$^\circ$ North at the central meridian (top row) and 30$^\circ$ North and 30$^\circ$ West (bottom row). The side buffer zones are discarded everywhere, while the height of the top buffer zone is shown by the dashed vertical line.
 }
\end{figure}

Figure~\ref{f_bifrost_385_NLFFF_deltaVSheight_B_vecB_offCenter} shows the same metrics for two off-center locations of the data cube at the solar surface for bin\,=\,9 and the additional constraints, which include $B_\|$ (instead of $B_z$ in the top-view geometry) and the combination of the $B_\|$ and $|\textbf{B}|$ at the chromospheric level. The trends in this figure are similar to those for the top view, although the metrics slightly change quantitatively.

\subsection{Field Lines and Magnetic Connectivity}

The ability of a given NLFFF reconstruction tool to truthfully reproduce the magnetic field lines is of primary importance for many reasons. Qualitatively, a subset of the field lines is often used for validation of the extrapolation cube via  visual comparison of those field lines with bright EUV loops. It is also essential in modeling magnetic connectivity in solar flares; for example, in case of footpoints seen in hard X-ray emission from flares, these footpoints must be connected by a valid field line. Thus, the ability of a given model to closely reproduce this field line is fundamentally needed to develop a realistic 3D model of the flare \citep{Fl_etal_2011, Nita_etal_2015, 2018ApJ...852...32K}. In addition, tracing all available field lines in a data cube is now employed for ``dressing'' magnetic skeletons with a realistic thermal structure; thus, the truthful reproduction of the field lines is critical for building a realistic thermal model of active regions.

To check how truthfully the magnetic field lines are reproduced in our reconstructed data cubes (vs the original data cube),  we developed an algorithm that calculates magnetic field lines in the reconstructed and the original magnetic cubes starting from a given point at the base boundary and computes the largest spatial deviation between those field lines. Magnetic field lines was constructed using Runge-Kutta-Fehlberg method of
4(5) order \citep{Press:2007:NRE:1403886}. For our tests we produced all field lines that start at all lower boundary voxels with the magnetic field strength exceeding 30~G. {This way, each field line in the original magnetic cube has its corresponding field line in the reconstructed cube, which are quantitatively compared between each other as follows.}
%For each pair of the field lines starting from the same photospheric voxel, we compute the largest distance between them as follows.
For every point on the model field line, we select the closest point on the corresponding reconstructed field line {and compute distance between them; thus, each point on the original field line is characterized by some distance from the reconstructed field line. Then, we pick the largest of the distances, which is, by construction, is the maximum deviation between these two field lines}.

\begin{figure}\centering
	
	\includegraphics[width=0.49\linewidth]{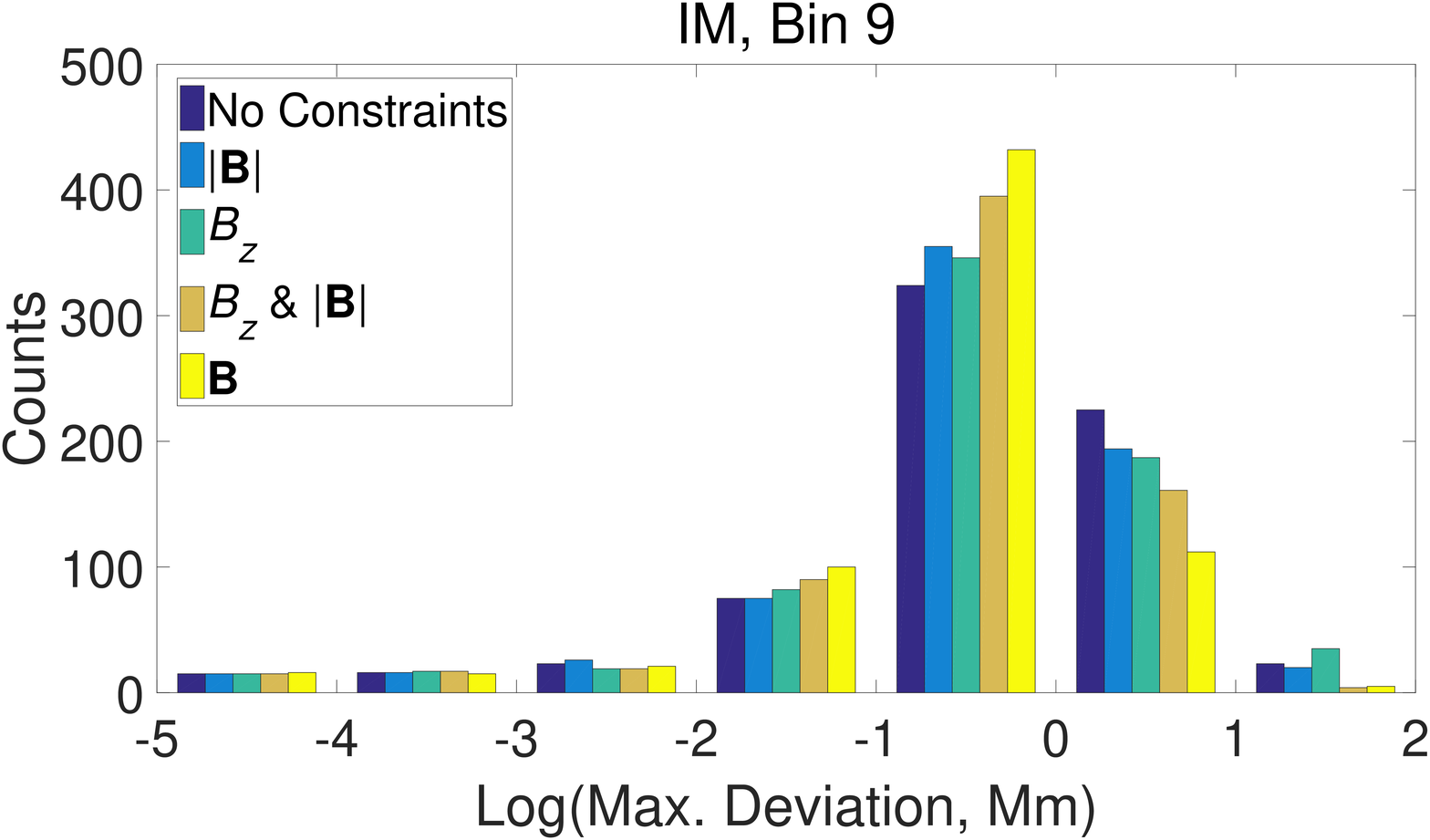}
	\includegraphics[width=0.49\linewidth]{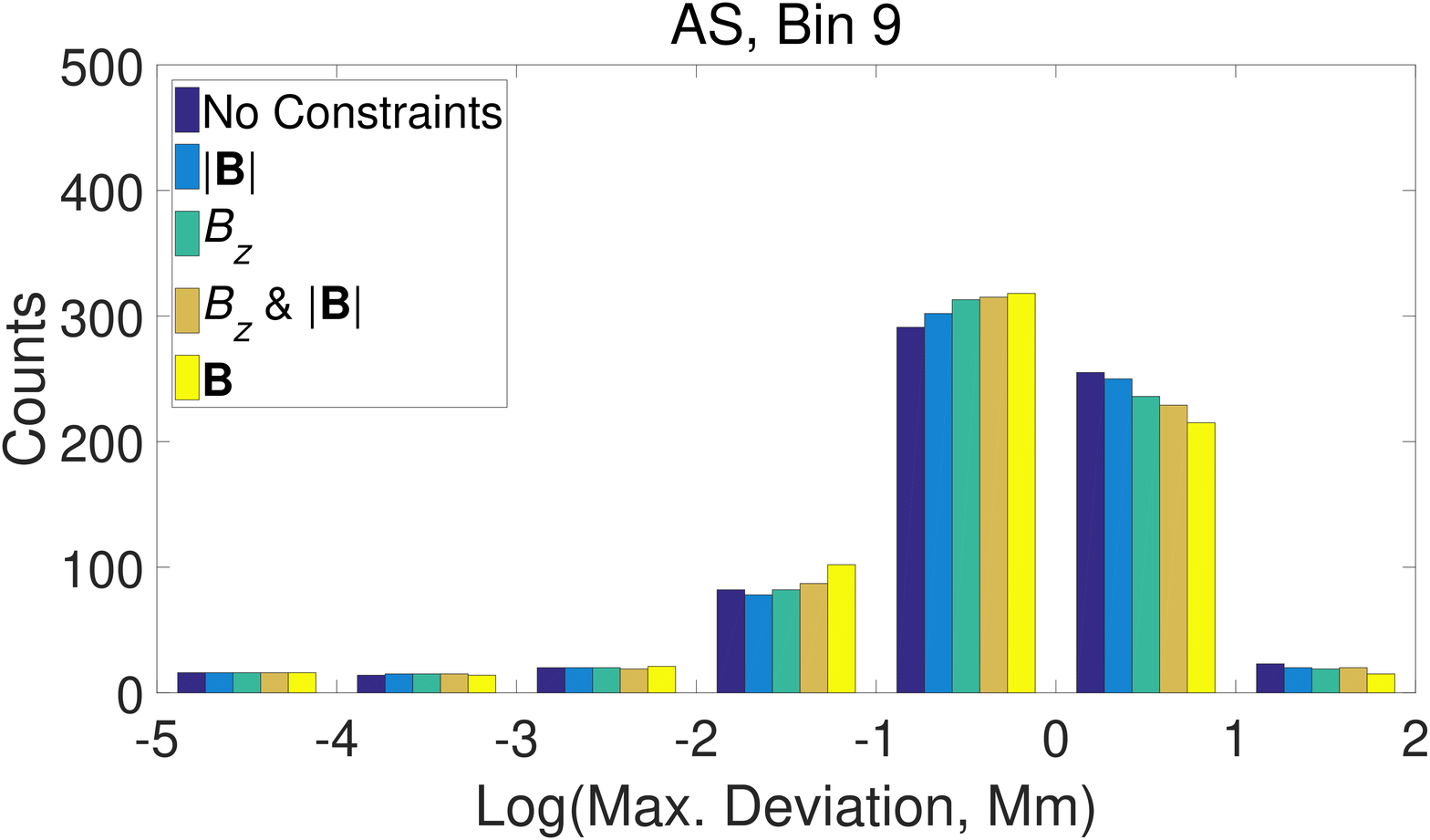}
	\includegraphics[width=0.49\linewidth]{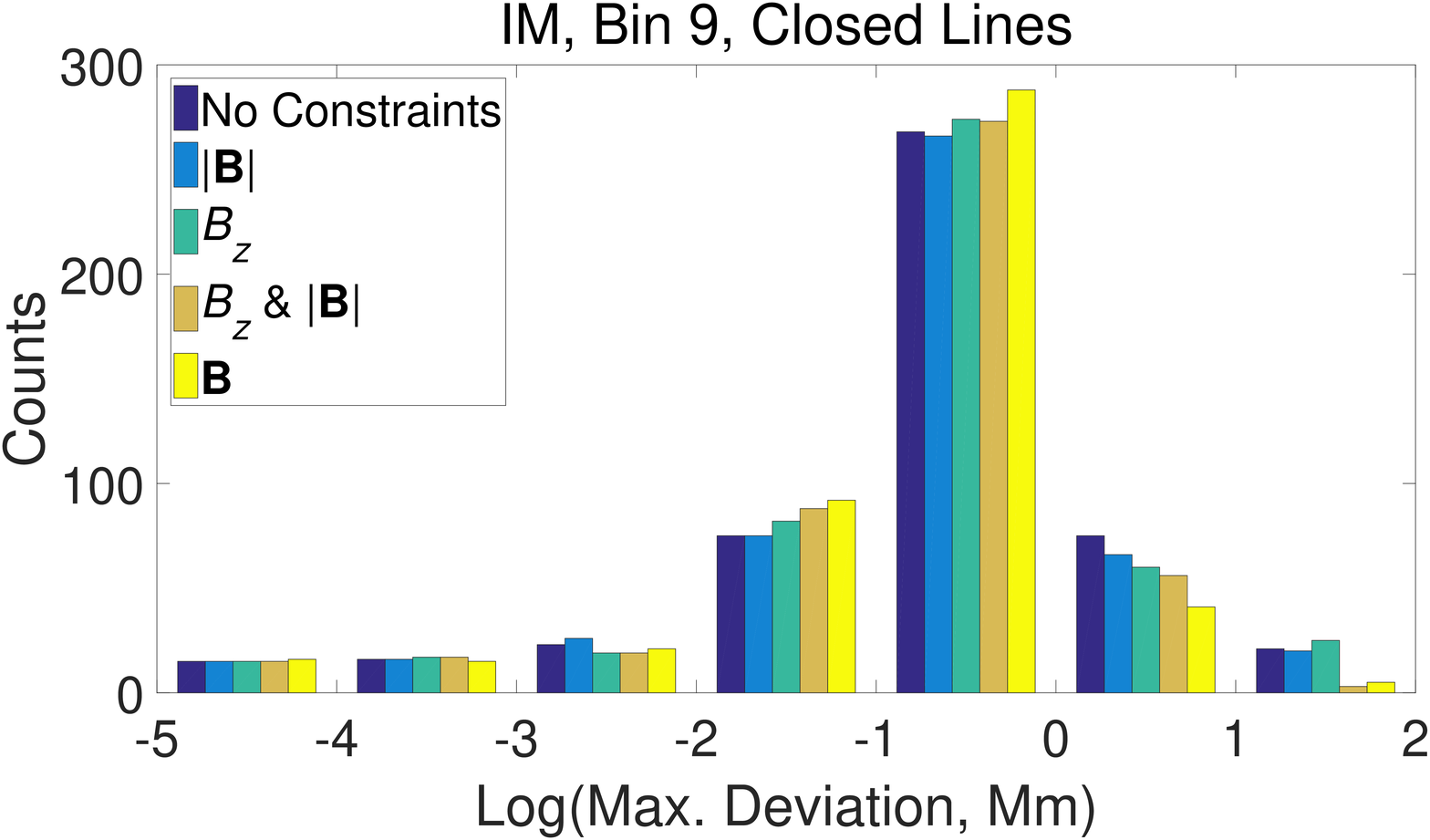}
	\includegraphics[width=0.49\linewidth]{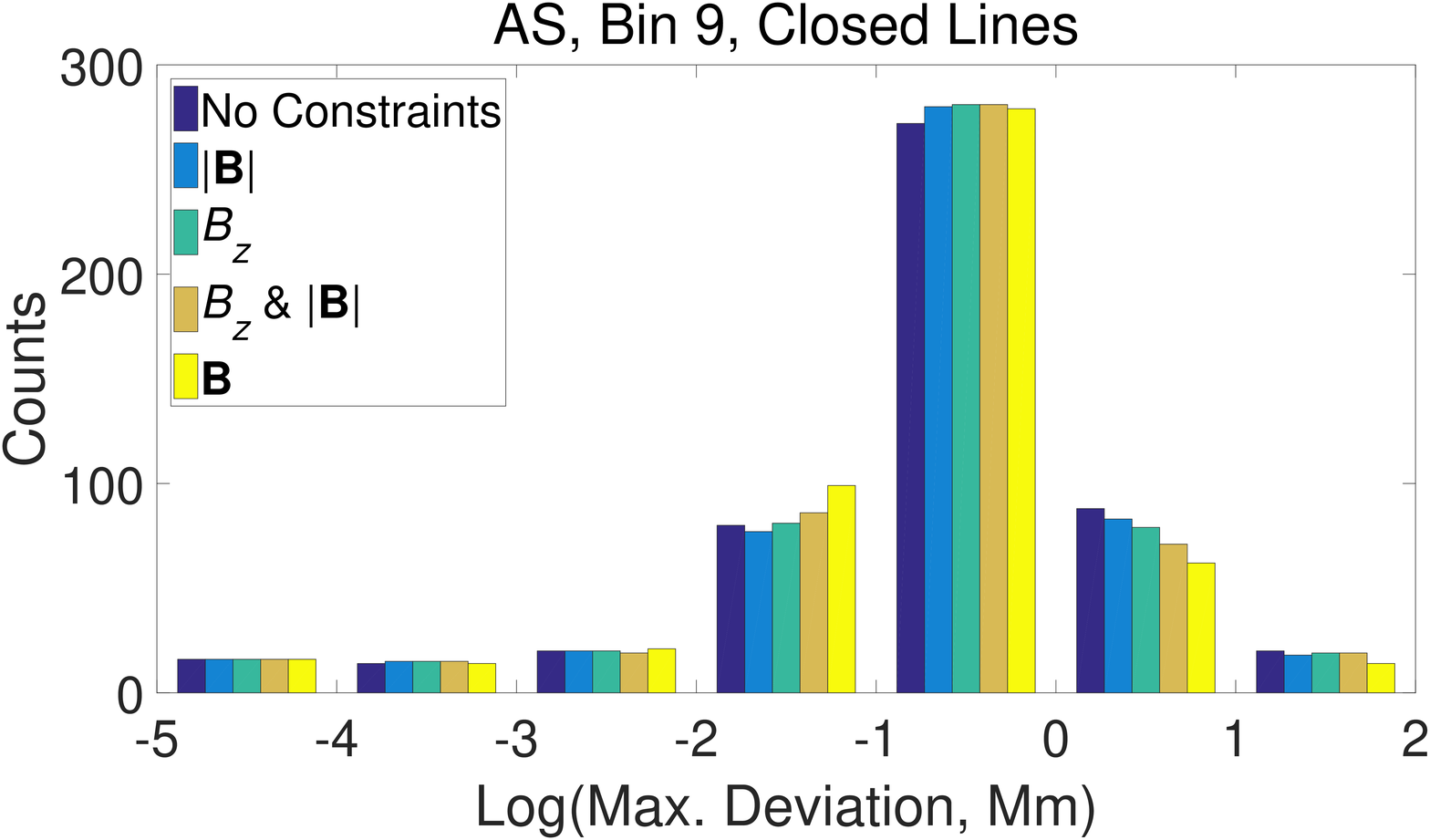}
	\includegraphics[width=0.49\linewidth]{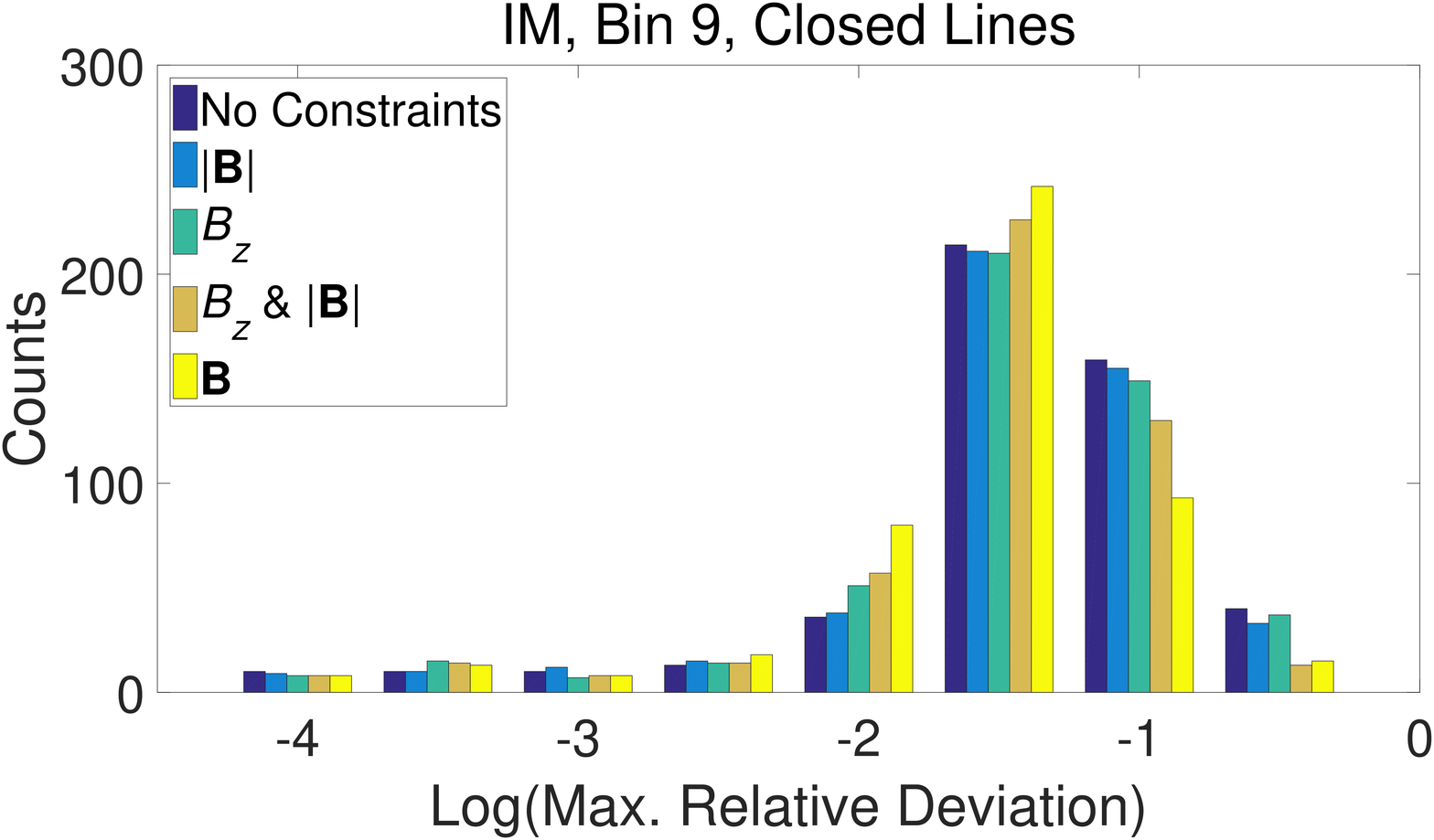}
	\includegraphics[width=0.49\linewidth]{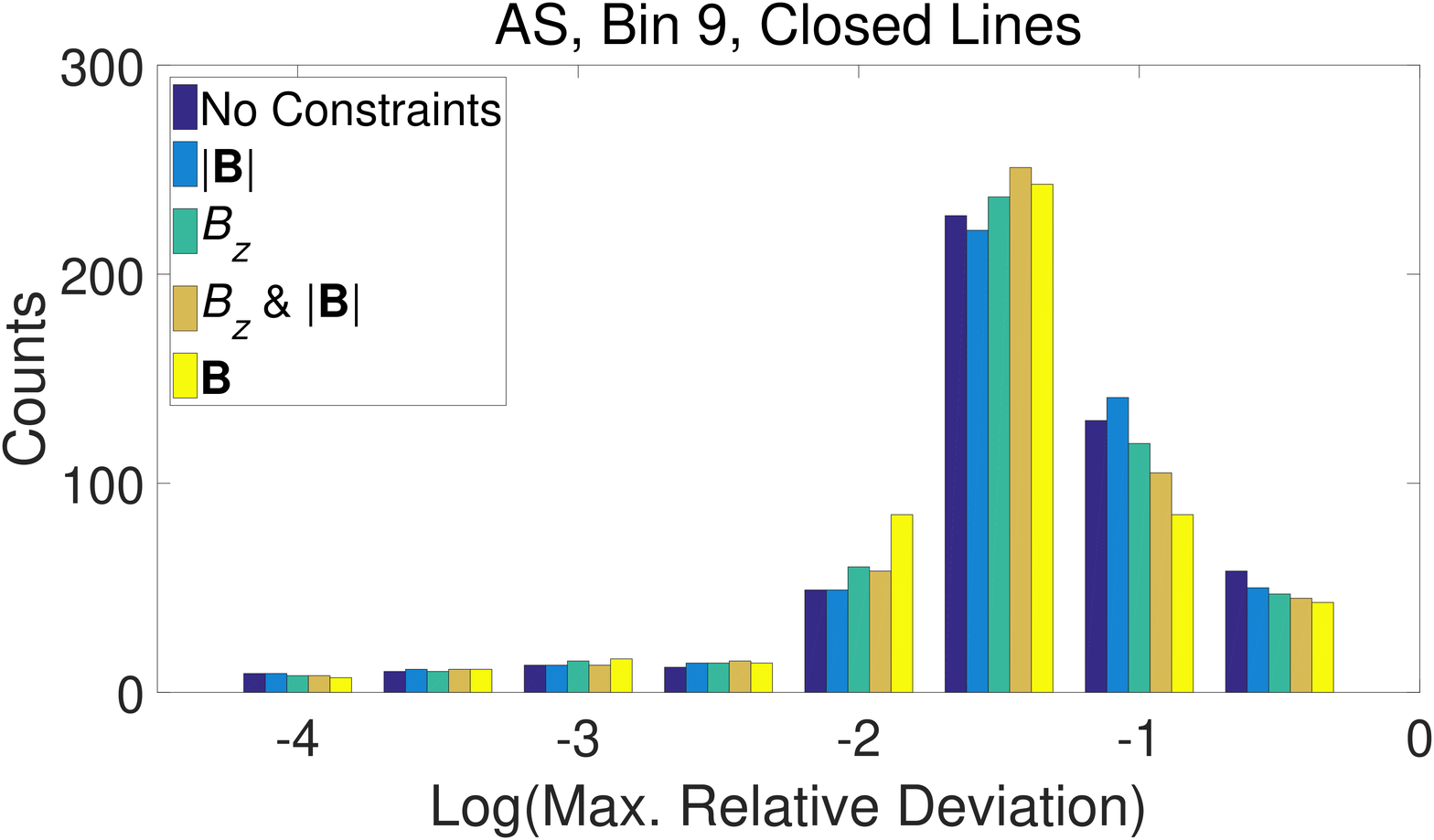}
	
	\caption{\label{fig:Hist_Bin9_Log}
		Histograms of magnetic field line deviations (bin\,=\,9). The first row: field lines built for all base pixels; the second row: a subset of those lines, which are closed in the original model data cube; the third row: relative deviations, $D/L$, where $L$ is the length of the field line in the extrapolated data cube, for the closed field lines.
	}
\end{figure}

Figure~\ref{fig:Hist_Bin9_Log}  shows the histograms of the maximum deviation between model magnetic field lines and reconstructed ones for various reconstruction methods and constrains. Most field lines show the maximum deviations of the order of 1\ Mm or less, mostly in the range 0.1\,--\,1~Mm. It is interesting that even though having information about the absolute value resulted in stronger improvement of the residual metrics than having the LOS component, the connectivity is somewhat better recovered for the LOS component case. All algorithms tend to produce a more ``closed'' magnetic field than that in the original data cubes. Thus, we separately tested how well the connectivity of the closed field lines is reproduced and found that indeed the connectivity is better reproduced for the closed field lines compared with all field lines. This is not surprising given that the cases when the reconstruction returns a closed field line instead of the open one may result in large deviations between them. A tendency of improving the quality of the connectivity reconstruction with adding more chromospheric constraints is evident from Figure~\ref{fig:Hist_Bin9_Log}.
Statistics of maximum deviation between model and reconstructed lines for various bins (3, 6, and 9) are shown in Tables~\ref{Stat_Deviations_Table}\,--\,\ref{Stat_Deviations_Closed_Table} for all and only closed field lines, respectively. The median values improve from 0.5\,--\,0.7~Mm without constraints to 0.3\,--\,0.5~Mm with most constrains for all field lines, while from $\sim0.3$~Mm to $\sim0.2$~Mm for the closed field lines. The quality of the connectivity reconstruction improves for higher bin-factors; the IM and AS codes yields comparable results for the closed lines case.

\begin{figure}\centering
	
	\includegraphics[width=0.49\linewidth]{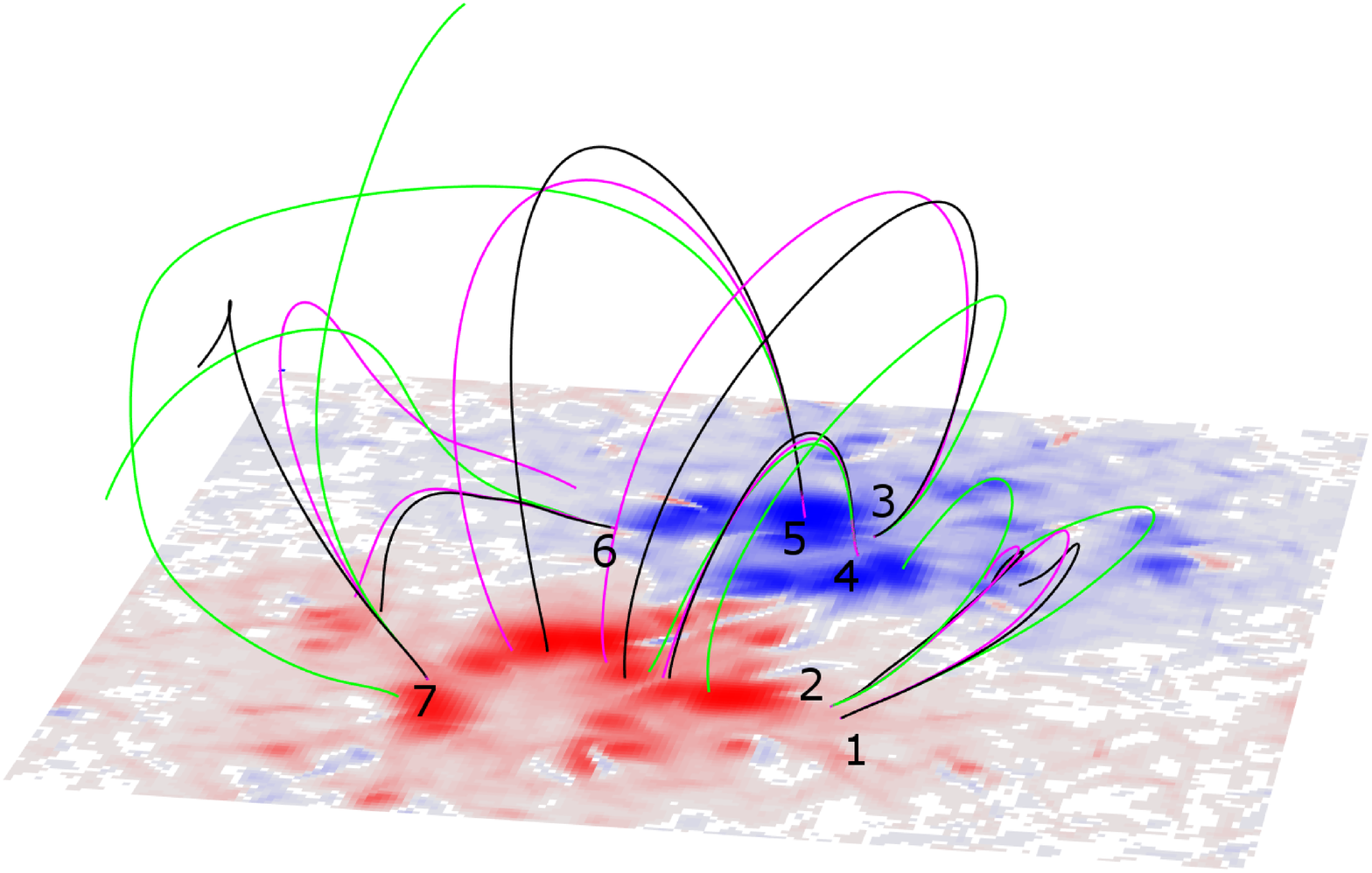}
	\includegraphics[width=0.49\linewidth]{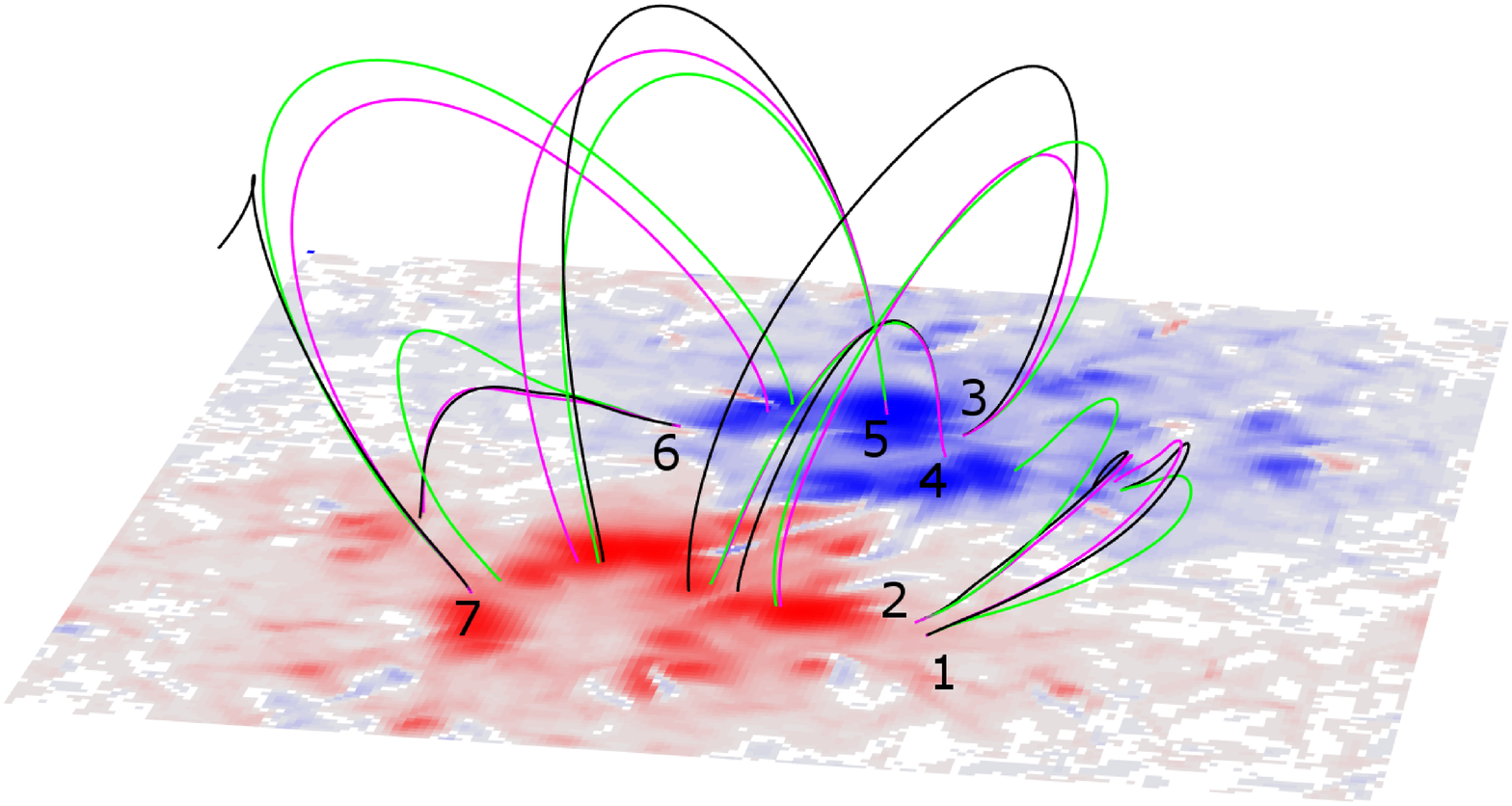}
	
	\caption{\label{fig:Front_Bin3}
		Examples of some magnetic field lines. Left\,--\,IM reconstruction, right\,--\,AS one. The model lines are shown in  black, the lines reconstructed without constrains  are in green,  the lines reconstructed with $B_z$ \& $\left| \mathbf{B} \right|$ constrains are in magenta. The animated version of these plots, showing the field lines from various perspectives, is available on-line.
	}
\end{figure}

{To scale our results obtained for a given size of the box and given distances between the strong field areas to an arbitrary active region, we also computed relative deviations, $D/L$, where $L$ is the length of the field line in the extrapolated data cube. In most cases the relative deviation is of the order of 10\% or even less, which implies that in most cases the connectivity is reproduced reasonably well. However, there is a noticeable subset of field lines with a larger deviation, which has to be taken into account in analysis of real data.}

Animated Figure~\ref{fig:Front_Bin3} displays a few representative field lines intended to illustrate, though not in a statistically meaningful way, those various cases, when the magnetic connectivity is affected by adding chromospheric constraints. One can see that in most cases there is a noticeable improvement in the magnetic connectivity due to adding the chromospheric constraints {(lines 1, 2, 6 in both reconstruction methods, lines 4, 5 for IM method)}, although in some cases the connectivity in the data cube obtained with additional constraints can occasionally become worse {(say, line 5 in AS method). Line 7 shows the tendency to produce a more ``closed'' magnetic field than that in the original data cubes}.

\subsection{Maps of Electric Currents}

To investigate how additional internal constraints affect distribution of the electric current density, we calculated electric current map for the re-binned Bifrost model and compared it with the maps obtained from the extrapolated data cubes.
The electric current maps are computed by integration of the 3D distribution of the electric current over $z$-axis: % {\gf [GF: chromo- vs full height integration]}:
\begin{equation}\label{key}
I(x,y) \propto \int |\vec{\nabla} \times \vec{B}(x,y,z)|dz.
\end{equation}

\begin{figure}\centering
	
	\includegraphics[width=0.95\linewidth]{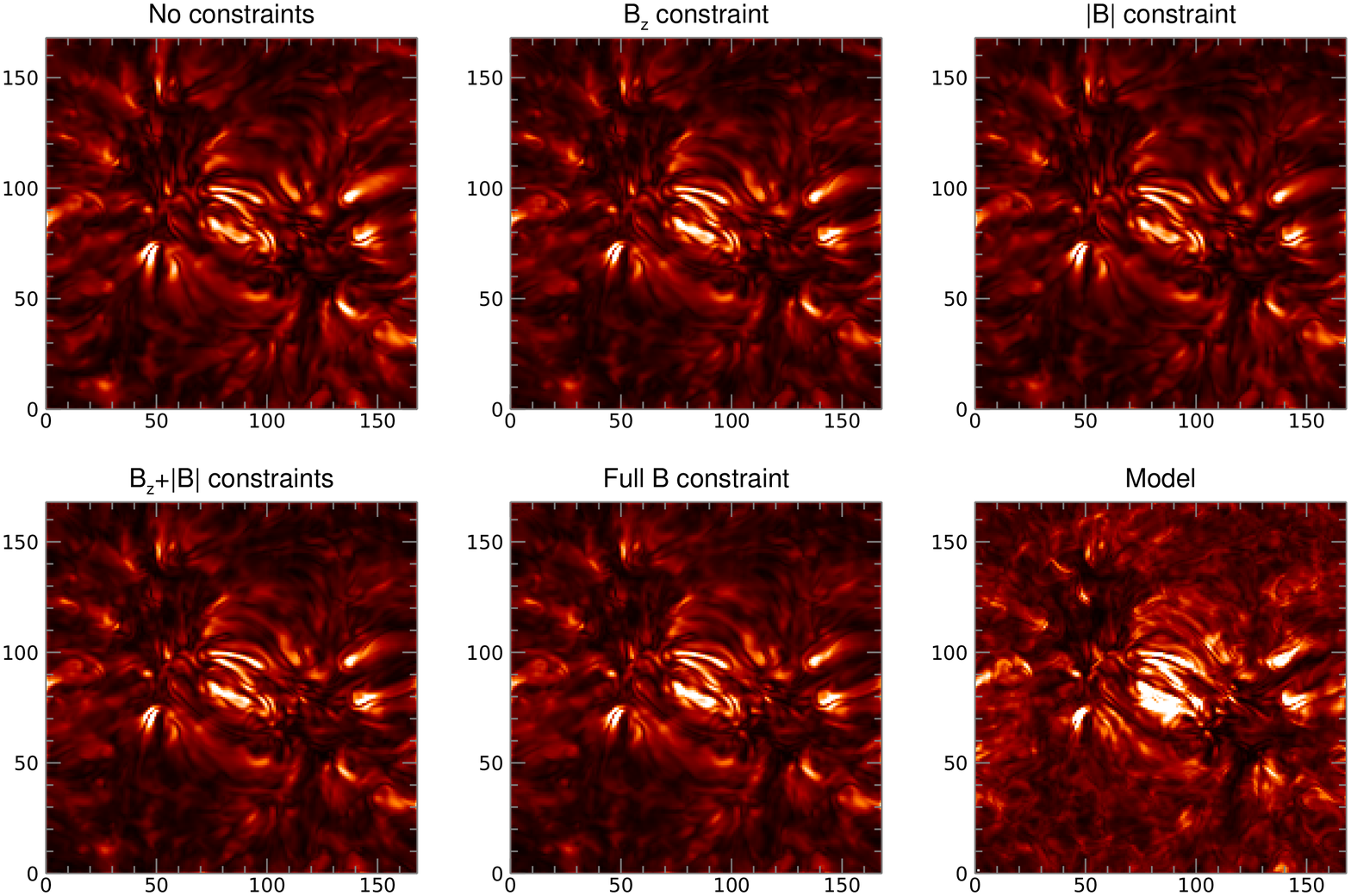}
	\includegraphics[width=0.95\linewidth]{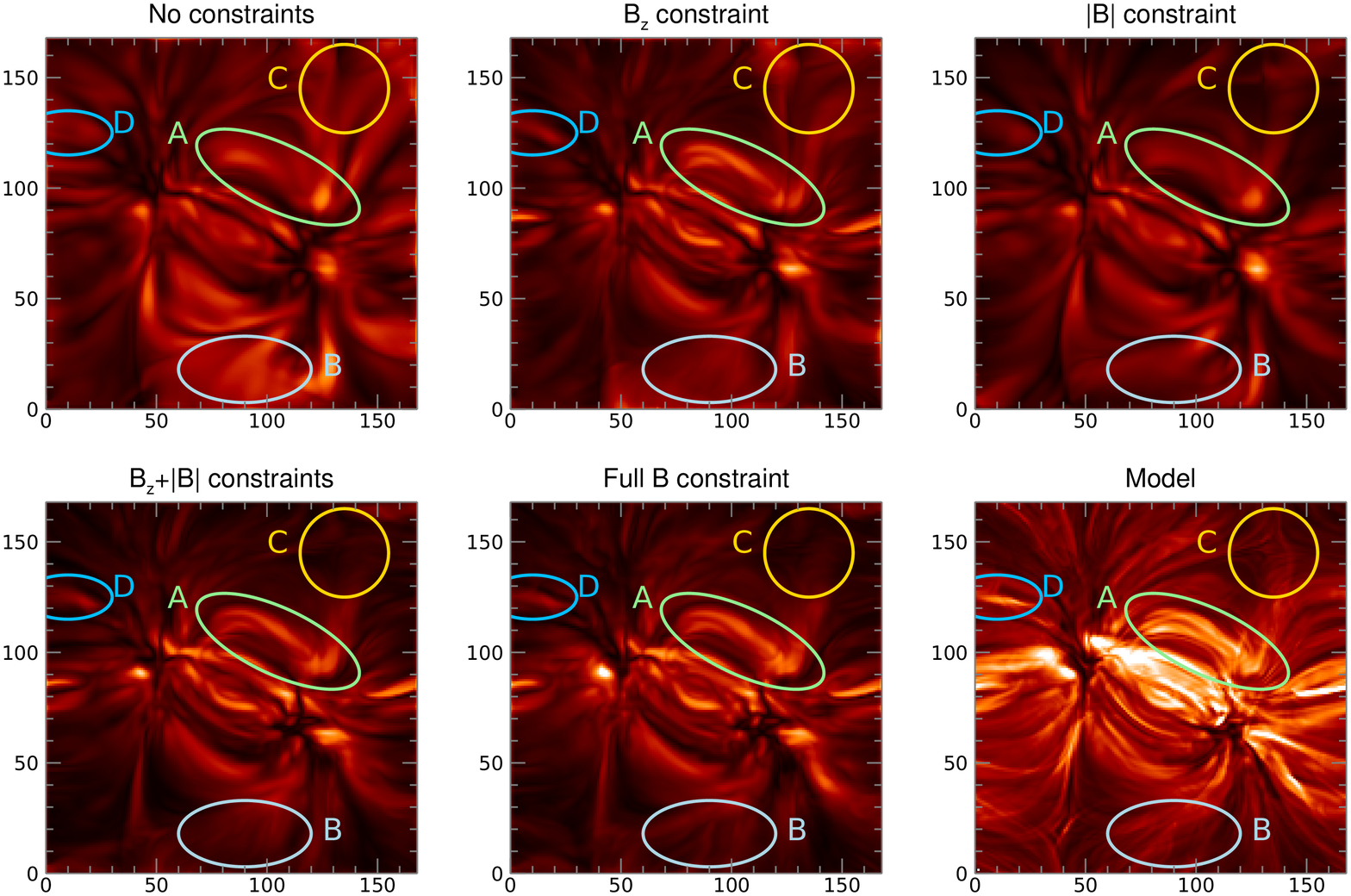}
	\caption{\label{fig:currents_IM}
		 Electric current maps calculated for IM extrapolations, bin\,=\,3, for the bottom layer between the photospheric boundary and the level from which the chromospheric constraints are taken (six top panels) and for the coronal volume above the chromospheric level (six bottom panels). Coloured ovals indicate features in the electric current distribution that have been improved after applying additional inner constraints: enhanced current in the core region of the box (A), false current near one of the boundaries (B), X-shaped geometrical feature (C), and the electric current enhancement near the left boundary (D). The same brightness scale is applied to all panels.{ The animated version of this figure shows how the electric current changes with height layer-by-layer. The brightness scale is kept fixed at all layers. However, to make the spatial structure visible at all layers, we show the brightness to the power of 0.2.}
	}
\end{figure}

\begin{figure}\centering
	
	\includegraphics[width=0.95\linewidth]{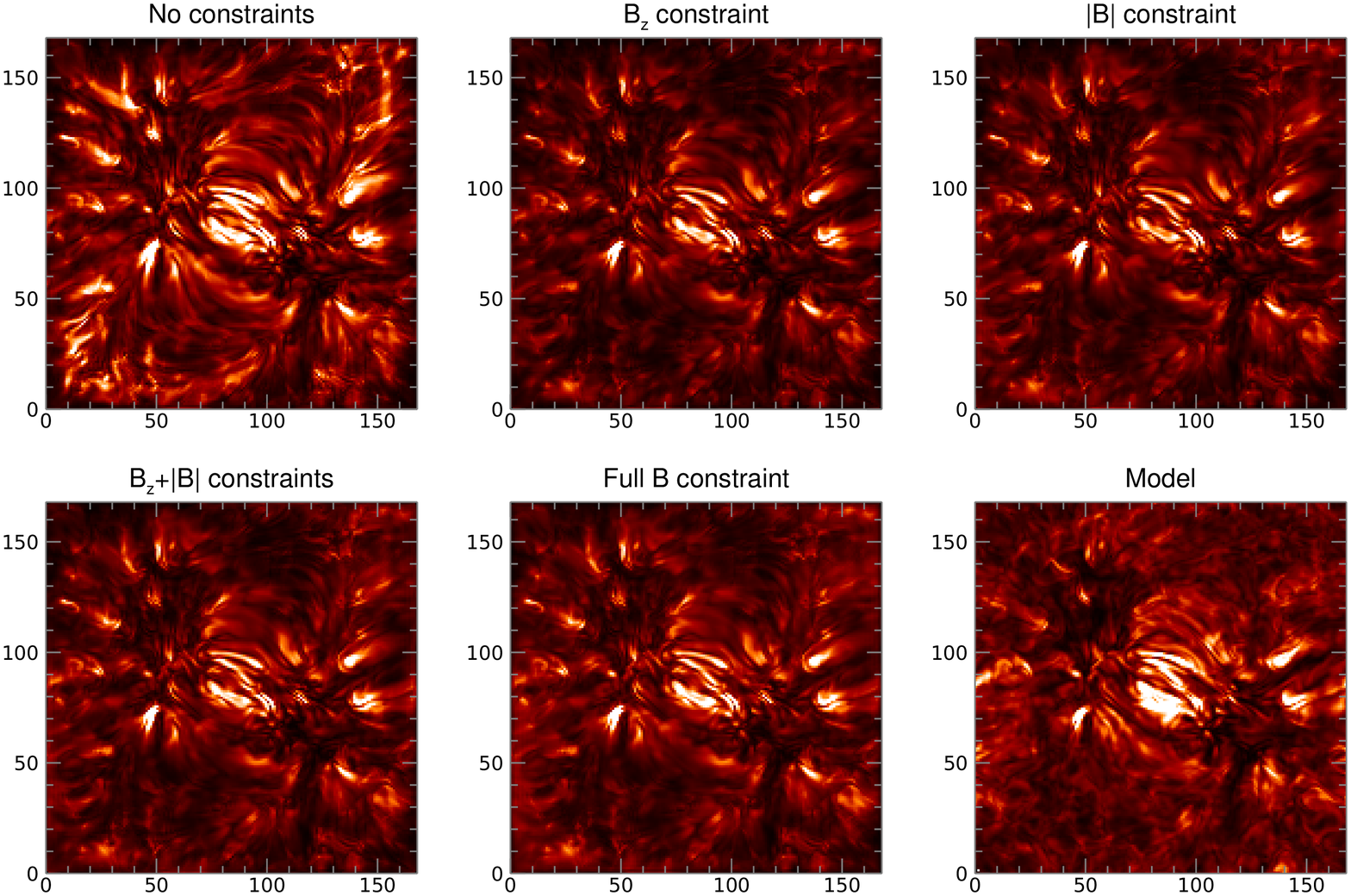}
	\includegraphics[width=0.95\linewidth]{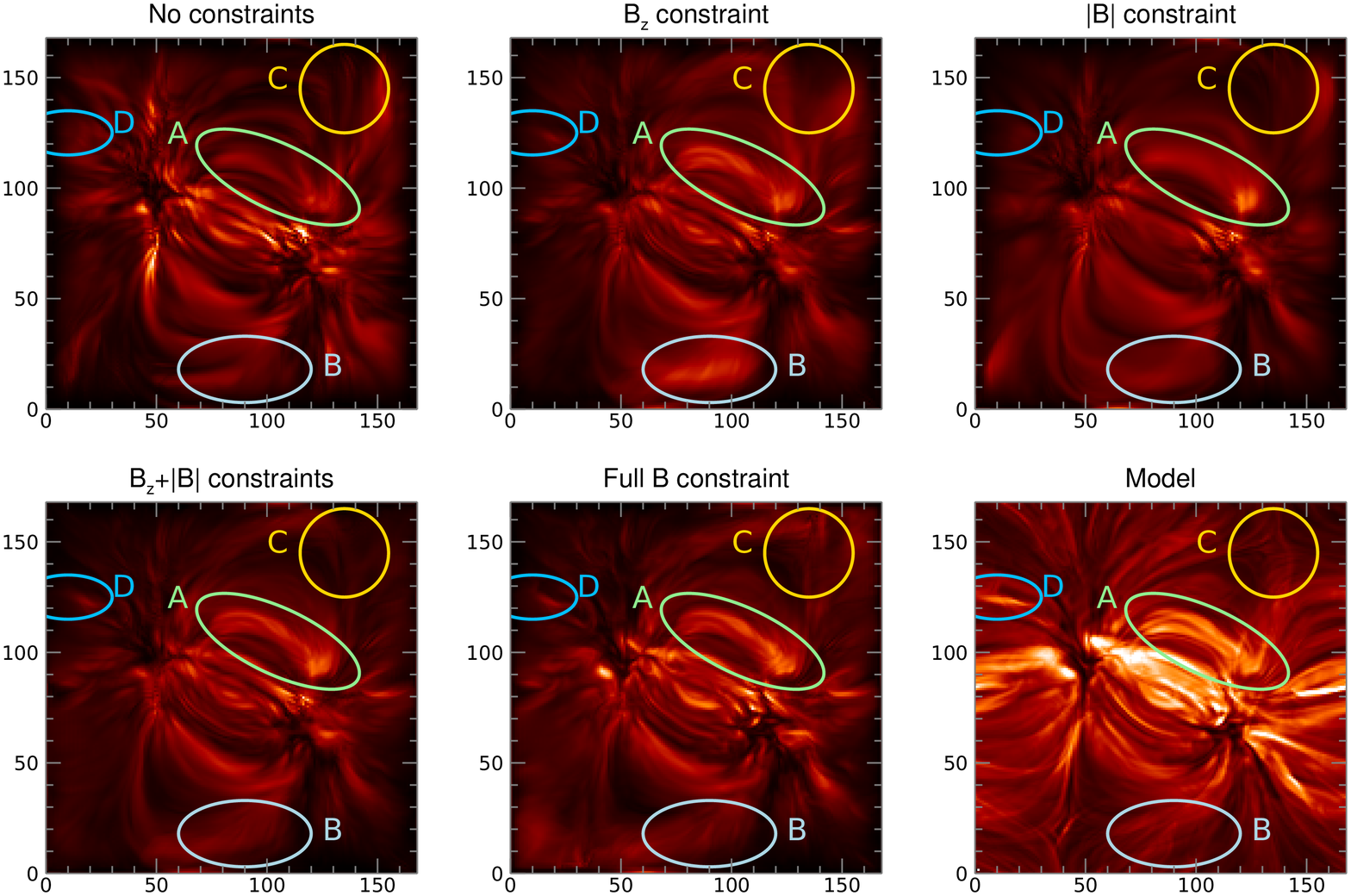}
	\caption{\label{fig:currents_AS}
		Electric current maps calculated for AS extrapolations, bin\,=\,3, for the bottom layer between the photospheric boundary and the level from which the chromospheric constraints are taken (six top panels) and for the coronal volume above the chromospheric level (six bottom panels). Coloured ovals indicate features in the current distribution that have been improved after applying additional inner constraints: enhanced current in the core region of the box (A), false current near one of the boundaries (B), X-shaped geometrical feature (C), and electric current enhancement near the left boundary (D). The same brightness scale is applied to all panels. { The animated version of this figure shows how the electric current changes with height layer-by-layer. The brightness scale is kept fixed at all layers. However, to make the spatial structure visible at all layers, we show the brightness to the power of 0.2.}
	}
\end{figure}

The results obtained for IM and AS extrapolations are given in Figures \ref{fig:currents_IM} and \ref{fig:currents_AS}, respectively, separately---for the layer between the photospheric and chromospheric levels (six top panels) and for the coronal volume above the chromospheric level (six bottom panels).
In the bottom layer the results of extrapolations with different combinations of additional constraints look morphologically similar to each other. This is likely because the electric currents are becoming weaker at the higher heights and so the upper part of this layer, which is stronger affected by the chromospheric constraints, gives a relatively minor contribution to the overall distribution of the electric current. However, a careful look on the maps reveals that small details of the electric current distribution are better reproduced, when chromospheric constraints are taken into account.
The electric current maps look surprisingly similar to the electric current map computed from the original model, even though the magnetic field is clearly non-force-free in this layer. AS code tends to amplify the electric currents close to the boundary buffer zone, especially when no chromospheric constraints are employed, which is an artifact of the assumption that the magnetic field approaches a potential field solution at the side boundaries.

Overall comparison in the coronal volume reveals that every additional constraint makes  the electric current density distribution closer to the model one.
The best correspondence is achieved when the full magnetic field vector is constrained at the chromospheric level.
To investigate the improvement  caused by  additional constrains in details, we selected 4 distinct features  visible in the current maps and compared their appearance in different cases.
These features are marked by coloured ovals in Figures \ref{fig:currents_IM} and \ref{fig:currents_AS}  and labelled with letters A\,--\,D.

\begin{figure}\centering

	\includegraphics[width=0.95\linewidth]{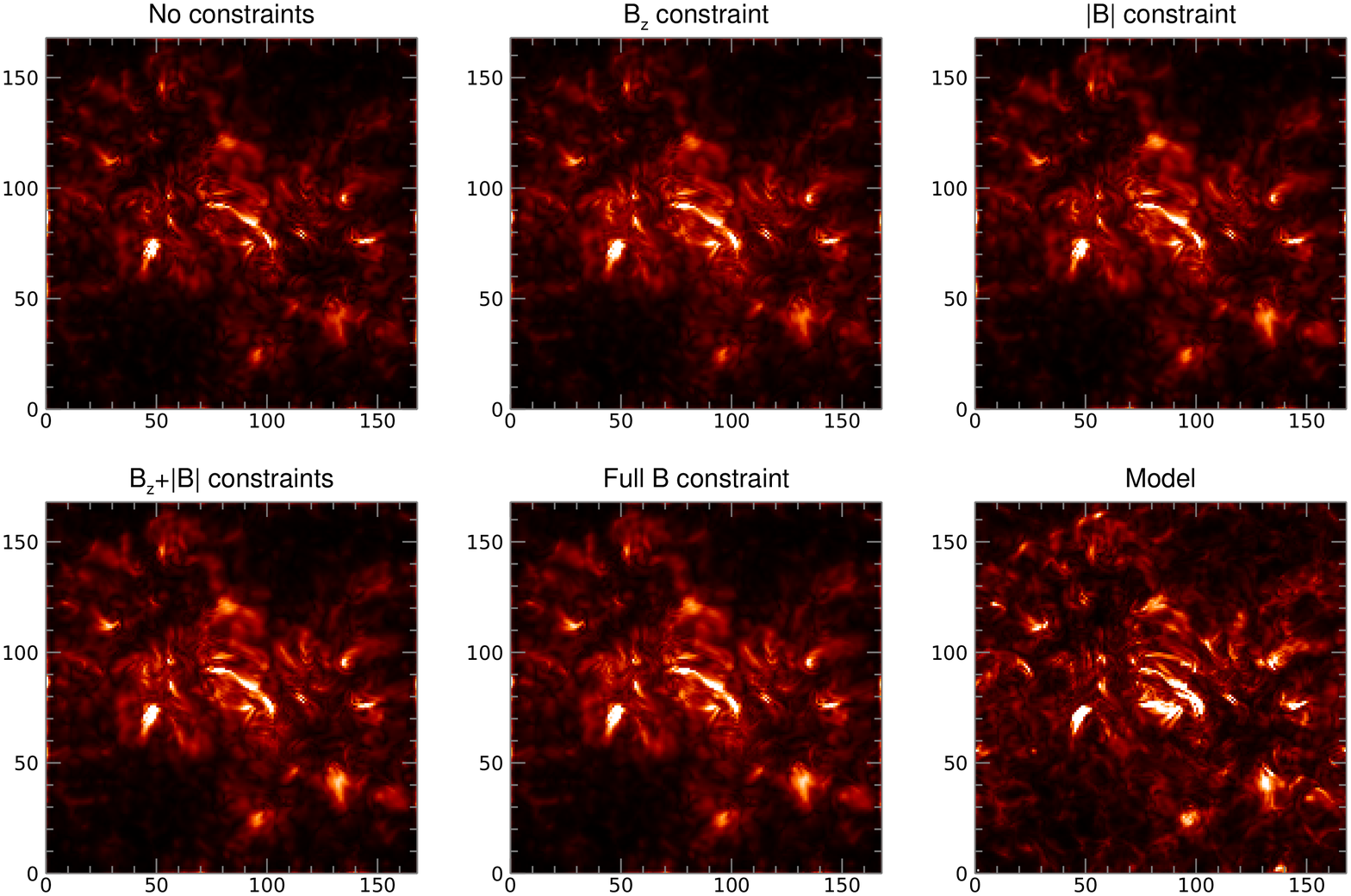}
	\includegraphics[width=0.95\linewidth]{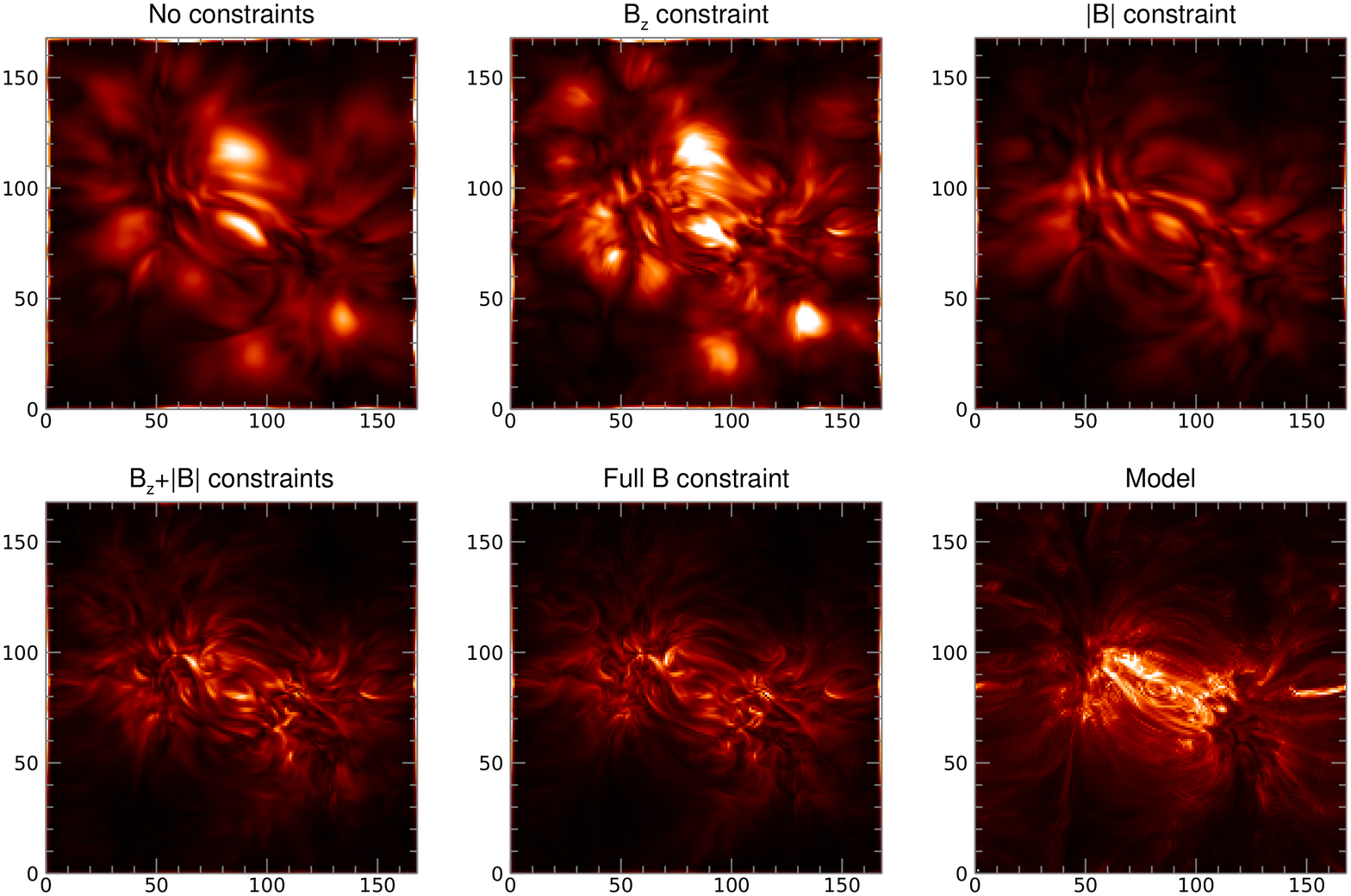}
	\caption{\label{fig:force_IM}
		Lorentz force maps calculated for IM extrapolations, bin\,=\,3, for the bottom layer between the photospheric boundary and the level from which the chromospheric constraints are taken (six top panels) and for the coronal volume above the chromospheric level (six bottom panels).{ The same brightness scale is applied to all panels corresponding to the same atmospheric layer, although the brightness scales for the chromosphere and the corona are different from each other. The animated version of this figure shows how the Lorentz force changes with height layer-by-layer. The brightness scale is kept fixed at all layers. However, to make the spatial structure visible at all layers, we show the brightness to the power of 0.2.}
	}
\end{figure}

\begin{figure}\centering

	\includegraphics[width=0.95\linewidth]{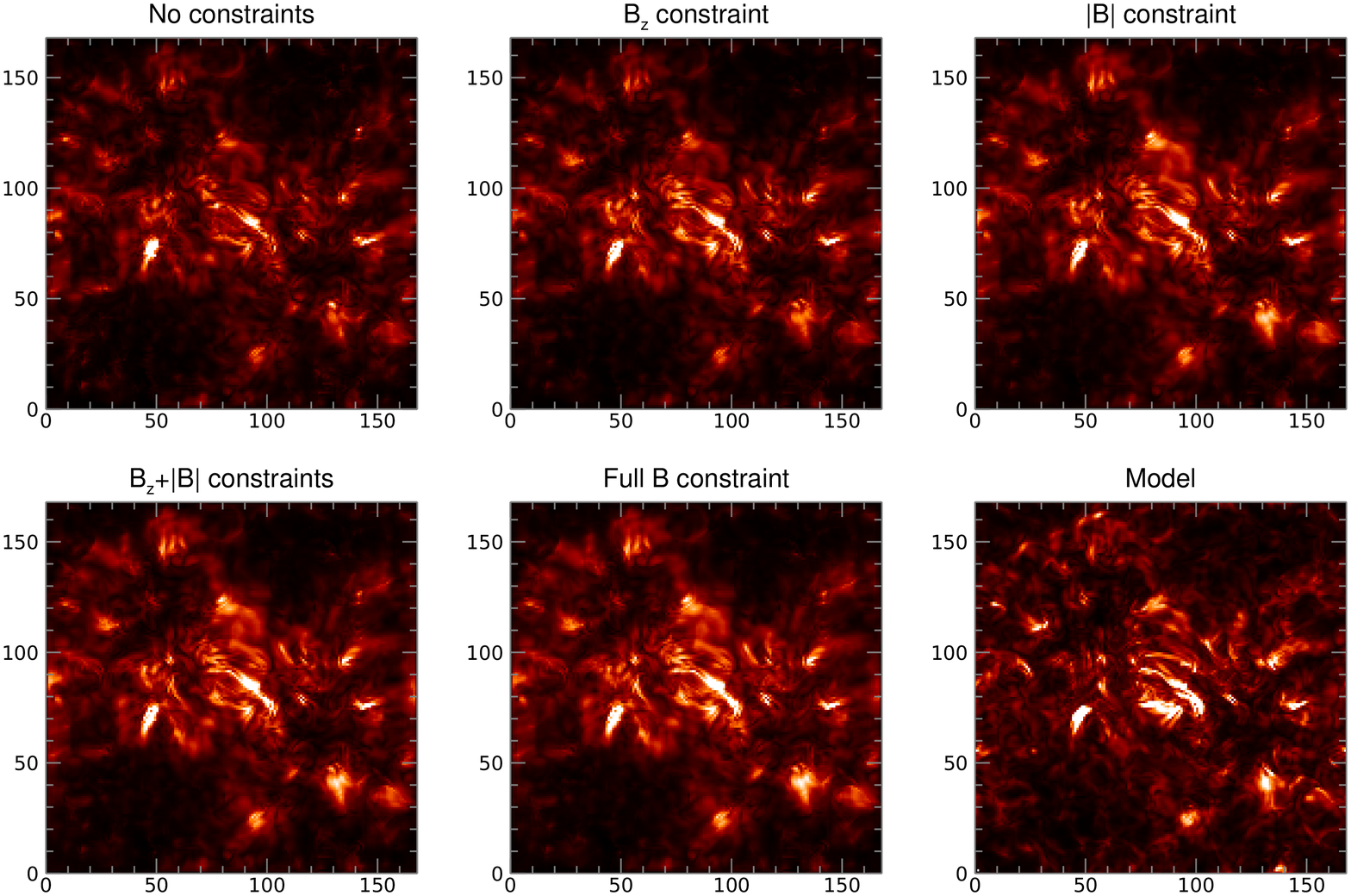}
	\includegraphics[width=0.95\linewidth]{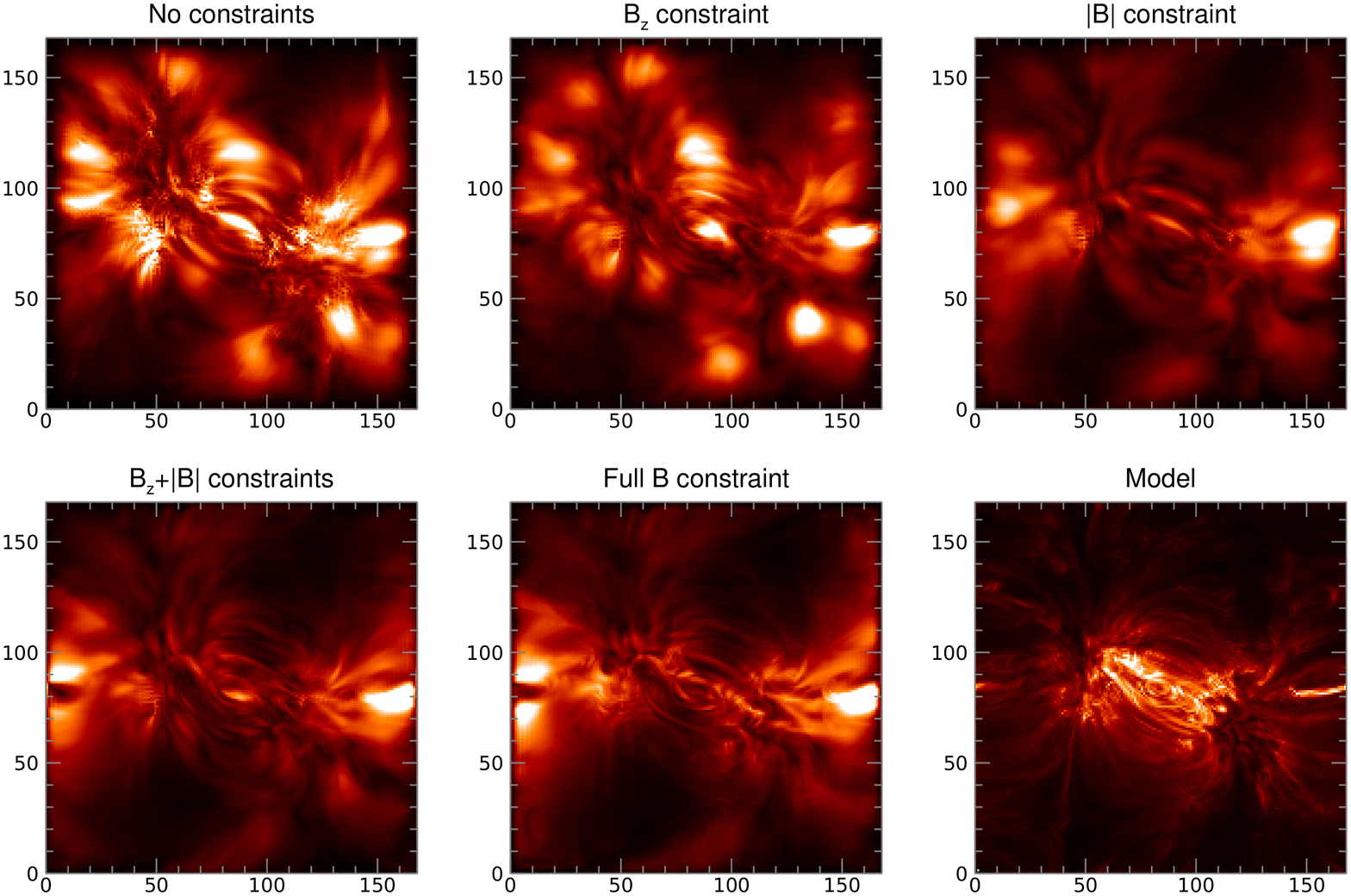}
	\caption{\label{fig:force_AS}
		Lorentz force maps calculated for AS extrapolations, bin\,=\,3, for the bottom layer between the photospheric boundary and the level from which the chromospheric constraints are taken (six top panels) and for the coronal volume above the chromospheric level (six bottom panels). {The same brightness scale is applied to all panels corresponding to the same atmospheric layer, although the brightness scales for the chromosphere and the corona are different from each other. The animated version of this figure shows how the Lorentz force changes with height layer-by-layer. The brightness scale is kept fixed at all layers. However, to make the spatial structure visible at all layers, we show the brightness to the power of 0.2.}
	}
\end{figure}

Significant enhancements (A and D) of the electric current density in the map calculated from the Bifrost model are much less prominent in both NLFFF extrapolations without any internal constraint.
Fixing the $B_z$ component at the chromospheric level makes these features more prominent and highlights their fine structure, while constraining the absolute value of the magnetic field {does not improve those features too much.}
This behaviour is the same for both IM and AS extrapolations.

Near the front boundary ($y=0$) of the box, there is area B with an  artificial electric current enhancement that is not present in the original model.
This artefact is most prominent  in the IM extrapolation without internal constraints and in the AS extrapolation with the $B_z$ constraint.
Thus, $B_z$ constraint does not remove this artificial current. Moreover, it can make it even more prominent (AS extrapolation case).
Adding a constraint to $|\textbf{B}|$ allows to get rid of this artificial enhancement and to make the electric current distribution in region B closer to the model one.

In the right top corner of the model map, there is an X-shaped feature.
Both methods are incapable of reproducing this feature without applying additional constraints.
IM extrapolation reveals this feature only after constraining both $B_z$ and $|\textbf{B}|$.
In AS extrapolation, this structure appears only after fixing full magnetic field vector.
The AS method is less accurate in this case probably because this structure is close to the boundary of the box, which is a fixed potential field in AS implementation, while the IM code varies it.
This effect is also manifested in the form of dark areas at the boundaries following the buffer zones, where the magnetic field approaches the current-free, potential field at side boundaries (e.g see feature D in the electric current maps for  AS extrapolation).

Our analysis of the electric current maps revealed that additional constraints improve qualitative  agreement between the NLFFF extrapolation and model.
The $B_z$ constraint improves the appearance of the electric current enhancements making them more prominent and revealing their fine structure.
The main benefit of applying internal constraints to the absolute value of the magnetic field is reducing artificial currents (e.g feature B in the electric current maps).
We also note that the current density in NLFFF extrapolation is noticeably lower than in the reference model.

\subsection{Maps of the Lorentz Force}

{\gf Note that although the minimization of the Lorentz force term in  functional described by Eqn~(\ref{Eq_nlfff_func_chr}) drives the solution towards a more force-free configuration, the use of non-force-free bottom boundary conditions and extra constraints in the volume  \textit{is expected to} result in a residual Lorentz force in both the chromospheric layer and the coronal volume. To assess how well this residual Lorentz matches that in the original model (recall, that the original model deviates from the force-free state even in the coronal part of the model data cube, see Table~1 in Paper~I for more detail) we employ the Lorentz force maps.

 Figures \ref{fig:force_IM} and \ref{fig:force_AS} demonstrate chromospheric and coronal Lorentz force maps for the IM and AS extrapolations respectively.
	Indeed, we see significant reduction of the Lorentz force  in the coronal part of the model due to the additional chromospheric constraints.
	It's remarkable, that the most significant improvement is achieved when the absolute value of the magnetic field is constrained.
	For the core region of the model, this behaviour  is demonstrated by both IM and AS method.
	However in the case of the AS datasets, there is no improvement in the  buffer zone near the side boundaries, where the magnetic field becomes potential.

Somewhat unexpected behaviour is observed in the lower layer between the photosphere and the chromosphere,  where we don't see any significant influence of the additional boundary conditions on the  distribution of the Lorentz force.
The most plausible explanation  is that the magnetic field in the lower layers of the datacube is mainly determined by the photospheric non-force free boundary conditions, and additional chromospheric constraints have  only a limited influence on the underlying layers.

}

\section{Discussion}

%- Summarize the study.

Here we have extended the earlier tested (Paper I) NLFFF optimization codes to situations when a subset of chromospheric data on the magnetic field is available in addition to the photospheric vector boundary condition. These new implementations of the extrapolation codes has been thoroughly tested using a series of magnetic data cubes derived in Paper I from the \textit{en024048{\_}hion} simulation \citep{2016A&A...585A...4C} obtained with the Bifrost code \citep{2011A&A...531A.154G}.

We have found that generally, even adding an incomplete set of the chromospheric data is capable of improving the NLFFF reconstruction. However, the effect of these additional constraints is different for the full optimization (IM code) or weighted optimization (AS code) algorithms. Specifically, when only a limited information is available, for example, the LOS component of the magnetic field or its absolute value, the AS code returns better results than the IM code. In contrast, when two components of the magnetic field (e.g., the LOS component and the absolute value) are available, the IM code performs better than the AS code. The reasons for that behavior are well understood: the IM code propagates the boundary conditions to the side and top boundary from the bottom of the data cubes. Thus, when reliable data at the almost force-free chromospheric level are available, then a more correct information is being propagated upwards, which helps to improve the reconstruction. In contrast, when these data are unavailable or incomplete, the weighted optimization approach with its buffer zone at the side and top boundaries turns out to be more appropriate.

The overall improvement in the magnetic field reconstructions with added chromospheric magnetic data is remarkable. It is especially significant in reconstructing the absolute value of the magnetic field, which is highly important for estimating the total and free magnetic energy in the data cube. An important finding is that adding only two components of the magnetic field at the chromospheric level, $B_\|$ and $|\textbf{B}|$, has almost the same positive effect as adding the full vector data. We believe, that this happens because having the two components along with equation $\nabla\cdot\textbf{B}=0$ is almost as complete as having the full vector data. {This finding is particularly important because it reveals that the \mw\ diagnostics of two magnetic field components at the TR from the GR and free-free emissions is, in fact, almost as complete as the full vector diagnostics. The latter might be available from infrared spectropolarimetry, but can suffer from errors in $\pi$-disambiguation of the transverse component of the magnetic field. The combination of $B_\|$ and $|\textbf{B}|$ is available from the infrared spectropolarimetry without any disambiguation, so the use of the $B_\|$ and $|\textbf{B}|$ combination rather than the full vector removes one source of errors.}

Adding the chromospheric constraints improves reconstruction of magnetic connectivity in the data cubes. The correctly reproduced connectivity in the coronal volume is of primary importance for all sorts of modeling performed using the NLFFF reconstructions, in particular, modeling of active regions \citep{Nita_etal_2018} and solar flares \citep{Nita_etal_2015, 2018ApJ...852...32K}. Finally, the reconstruction of electric currents in the modeling volume is also noticeably improved due to adding the chromospheric probing of the magnetic field.

We note that we have not considered all possible cases when a subset of chromospheric magnetic data might be available. For example, there can be cases when data on some component of the magnetic field (or the full vector) are only available at a portion of the chromospheric level, or at different chromospheric heights, or only within a certain range of the field amplitudes etc. There can be countless situations, which cannot all be considered at a systematic level. Instead, the modeling data cubes used in this study can be used to emulate those different situations as required by the actual data availability. For example, if the chromospheric data are available over a limited subarea of the active region of interest, one can produce a mask reproducing that subarea and apply this mask to the Bifrost modeling cubes to quantify the expected improvement of the reconstruction with the given subset of the available chromospheric constraints. A similar approach can be employed to account for the magnetic measurement errors.

%- Describe and discuss cases not accounted: many; e.g., (a) a subset of a layer is available; (b) data and height uncertainties; (c) data scattered over various layers; (d) info on field direction rather than values ... Given that so many combinations are possible, it would be more practical to make tests specifically for available combination of the data.

We also propose that other methods of reconstruction, which require a force-free bottom boundary to perform properly, are to be tested. Indeed, some other method might outperform the optimization methods employed here, provided that the force-free chromospheric boundary condition has been obtained by the optimization method.

%\begin{table}
%     \caption{ Divergence-free criterion $f \times 10^6$ (eq. 4 in 1st paper).
%     }
%     \label{divless}
%     \begin{tabular}{c | c c c c c c}

\begin{deluxetable*}{ c |c | c| c |c | c| c}
	\tabletypesize{ \small }
	\tablewidth{ 0pt }
	\tablecaption{ Divergence-free criterion $f \times 10^6$ (Eq.~\ref{IM_E02}). \label{T_divless} }
	\tablehead{
		\colhead{ Bin } & \colhead{ Model } & \colhead{ No Constr. } & \colhead{ $B _{z}$ } & \colhead{ $\left| \mathbf{B} \right|$ }  & \colhead{ $B _{z}$ \&  $\left| \mathbf{B} \right|$ } & \colhead{ $\mathbf{B}$ } \\
\colhead{  } & \colhead{  } & \colhead{ IM/AS } & \colhead{ IM/AS } & \colhead{ IM/AS }  & \colhead{ IM/AS } & \colhead{ IM/AS }
			}
	\startdata
           %\hline
%           Bin & Model & No constr. & $B_z$ & $\left| \mathbf{B} \right|$ & $B _{z}$ \&  $\left| \mathbf{B} \right|$ & $\mathbf{B}$ \\
%           \hline
           3 & 704  &  658/738 &  769/881 &  523/782 &  450/788 &  448/774 \\
           6 & 1587 & 1089/1251 & 1169/1429 & 889/1362 & 784/1399 & 802/1355 \\
           9 & 2440 & 1432/1611 & 1921/1857 & 1533/1828 & 1263/1893 & 1323/1825 \\
           \hline
     %\end{tabular}
%\end{table}
	\enddata
\end{deluxetable*}

\section{Conclusions}

Based on the tests performed here with the new NLFFF reconstruction codes that take into account additional chromospheric and coronal constraints along with the routinely available vector measurements at the photosphere, we conclude that the full use of all magnetic field measurements at force-free regions above active regions is highly helpful in improving the magnetic field modeling of active regions. We emphasise that even adding an incomplete set of data, that can include only one or two components of the magnetic field at the chromospheric level is already highly beneficial. This validates the effort of the research community to obtain such diagnostics from infrared, millimeter, and microwave measurements. We expect to see a dramatic increase of combining all such diagnostics in producing the magnetic models of active regions in near future.

\begin{deluxetable*}{ c c | c c c | c c c | c c c | c c c  | c c c}
	
	\tabletypesize{ \small }
	\tablewidth{ 0pt }
	\tablecaption{ Statistics of Maximum Deviation \label{Stat_Deviations_Table} }
	\tablehead{
		\colhead{ Bin } & \colhead{ Impl } & \multicolumn{3}{c}{ No Constr. } & \multicolumn{3}{c}{ $\left| \mathbf{B} \right|$ } & \multicolumn{3}{c}{ $B _{z}$ }  & \multicolumn{3}{c}{ $B _{z}$ \&  $\left| \mathbf{B} \right|$ } & \multicolumn{3}{c}{ $\mathbf{B}$ } \\
		& &
		\colhead{ $D_{0.5}$ } & \colhead{ $\overline{D}^*$ } & \colhead{ $D_{0.75}$ } &
		\colhead{ $D_{0.5}$ } & \colhead{ $\overline{D}^*$ } & \colhead{ $D_{0.75}$ } &
		\colhead{ $D_{0.5}$ } & \colhead{ $\overline{D}^*$ } & \colhead{ $D_{0.75}$ } &
		\colhead{ $D_{0.5}$ } & \colhead{ $\overline{D}^*$ } & \colhead{ $D_{0.75}$ } &
		\colhead{ $D_{0.5}$ } & \colhead{ $\overline{D}^*$ } & \colhead{ $D_{0.75}$ }
	}
	
	\startdata
	%    	\\
	\multirow{2}{*}{ 3 } & IM & 0.69 & 0.46 & 2.27 & 0.57 & 0.37 & 1.49 & 0.54 & 0.37 & 1.61 & 0.37 & 0.26 & 0.90 & 0.28 & 0.21 & 0.69 \\
	& AS & 0.67 & 0.43 & 2.15 & 0.70 & 0.44 & 2.00 & 0.61 & 0.41 & 1.82 & 0.59 & 0.39 & 1.64 & 0.49 & 0.34 & 1.43 \\
	\\
	\multirow{2}{*}{ 6 } & IM & 0.62 & 0.39 & 1.80 & 0.53 & 0.33 & 1.25 & 0.51 & 0.35 & 1.54 & 0.38 & 0.25 & 0.90 & 0.27 & 0.20 & 0.66 \\
	& AS & 0.60 & 0.40 & 2.13 & 0.58 & 0.41 & 1.97 & 0.49 & 0.37 & 1.88 & 0.48 & 0.35 & 1.65 & 0.40 & 0.31 & 1.43 \\
	\\
	\multirow{2}{*}{ 9 } & IM & 0.50 & 0.36 & 1.76 & 0.46 & 0.31 & 1.24 & 0.46 & 0.32 & 1.49 & 0.37 & 0.24 & 0.91 & 0.29 & 0.20 & 0.69 \\
	& AS & 0.53 & 0.39 & 2.41 & 0.49 & 0.38 & 2.29 & 0.43 & 0.35 & 2.06 & 0.42 & 0.34 & 1.81 & 0.36 & 0.30 & 1.62 \\
	%        \\
	\enddata
	
	\tablecomments{ Bin\,--\,is binning factor. Impl\,--\,is implementation of the optimization method.  $D_{0.5}$\,--\,is median, $\overline{D}^*=10^{\overline{\log(D)}}$, where $\overline{\log(D)}$\,--\,is mean value of decimal logarithm, $D_{0.75}$\,--\,is upper quartile of the distribution of maximum deviation in Mm. }

\end{deluxetable*}

\begin{deluxetable*}{ c c | c c c | c c c | c c c | c c c  | c c c}
	
	\tabletypesize{ \small }
	\tablewidth{ 0pt }
	\tablecaption{ Statistics of Maximum Deviation (Closed Lines) \label{Stat_Deviations_Closed_Table} }
	\tablehead{
		\colhead{ Bin } & \colhead{ Impl } & \multicolumn{3}{c}{ No Constr. } & \multicolumn{3}{c}{ $\left| \mathbf{B} \right|$ } & \multicolumn{3}{c}{ $B _{z}$ }  & \multicolumn{3}{c}{ $B _{z}$ \&  $\left| \mathbf{B} \right|$ } & \multicolumn{3}{c}{ $\mathbf{B}$ } \\
		& &
		\colhead{ $D_{0.5}$ } & \colhead{ $\overline{D}^*$ } & \colhead{ $D_{0.75}$ } &
		\colhead{ $D_{0.5}$ } & \colhead{ $\overline{D}^*$ } & \colhead{ $D_{0.75}$ } &
		\colhead{ $D_{0.5}$ } & \colhead{ $\overline{D}^*$ } & \colhead{ $D_{0.75}$ } &
		\colhead{ $D_{0.5}$ } & \colhead{ $\overline{D}^*$ } & \colhead{ $D_{0.75}$ } &
		\colhead{ $D_{0.5}$ } & \colhead{ $\overline{D}^*$ } & \colhead{ $D_{0.75}$ }
	}
	
	\startdata
	%    	\\
	\multirow{2}{*}{ 3 } & IM & 0.30 & 0.46 & 0.87 & 0.31 & 0.37 & 0.85 & 0.26 & 0.37 & 0.71 & 0.23 & 0.26 & 0.56 & 0.19 & 0.21 & 0.45 \\
	& AS & 0.31 & 0.43 & 1.06 & 0.39 & 0.44 & 1.07 & 0.36 & 0.41 & 1.00 & 0.33 & 0.39 & 0.89 & 0.26 & 0.34 & 0.71 \\
	\\
	\multirow{2}{*}{ 6 } & IM & 0.31 & 0.39 & 0.74 & 0.32 & 0.33 & 0.68 & 0.28 & 0.35 & 0.76 & 0.24 & 0.25 & 0.53 & 0.21 & 0.20 & 0.43 \\
    & AS & 0.28 & 0.40 & 0.80 & 0.30 & 0.41 & 0.78 & 0.26 & 0.37 & 0.69 & 0.25 & 0.35 & 0.66 & 0.21 & 0.31 & 0.52 \\
    \\
	\multirow{2}{*}{ 9 } & IM & 0.33 & 0.36 & 0.71 & 0.31 & 0.31 & 0.64 & 0.26 & 0.32 & 0.64 & 0.23 & 0.24 & 0.50 & 0.20 & 0.20 & 0.41 \\
	& AS & 0.29 & 0.39 & 0.83 & 0.30 & 0.38 & 0.71 & 0.23 & 0.35 & 0.64 & 0.24 & 0.34 & 0.56 & 0.19 & 0.30 & 0.48 \\
	%        \\
	\enddata
	
	\tablecomments{ Bin\,--\,is binning factor. Impl\,--\,is implementation of the optimization method.  $D_{0.5}$\,--\,is median, $\overline{D}^*=10^{\overline{\log(D)}}$, where $\overline{\log(D)}$\,--\,is mean value of decimal logarithm, $D_{0.75}$\,--\,is upper quartile of the distribution of maximum deviation in Mm. }
	
\end{deluxetable*}

\acknowledgments

This work was supported in part by NSF grant  AST-1820613
and NASA grants  %80NSSC18K0015, 
NNX16AL67G, and 80NSSC18K0667 to New Jersey
Institute of Technology (GF); by  RFBR  research projects 16-02-00254 (AS), 16-02-00749 (ML),  18-32-00540 mol$\_$a (IM), {17-52-18050 $\&$ the National Science Fund of Bulgaria under contract No. DNTS/Russia 01/6 (23-Jun-2017) (IM); by  Program No. 28 of the RAS Presidium (IM)}; by  budgetary funding of Basic Research program {II.16 (IM) and} II.16.3.2 “Non-stationary and wave processes in the solar atmosphere” (SA).

%\textbf{This study was supported by the Program of basic research of the RAS Presidium No. 9.}\sa{Who does acknowledge this?}

\bibliographystyle{apj}
\bibliography{NLFFF,fleishman,solar_radio,Zhitao_AR_refs}

\begin{deluxetable}{c|cccc|cccc|cccc|cccc}

\rotate
				
\tablecolumns{15}

\tablewidth{0pc}

\tabletypesize{\footnotesize}

\tablecaption{Normalized rms residual at a given level for bin-factor 9. \label{table_rms_res}}
				%
%=========Table 2===================				
\tablehead{
\multicolumn{1}{c|}{} & \multicolumn{4}{c|}{$\Delta_{rms}(B)$}& \multicolumn{4}{c|}{$\Delta_{rms}(B_x)$} & \multicolumn{4}{c|}{$\Delta_{rms}(B_y)$} & \multicolumn{4}{c}{$\Delta_{rms}(B_z)$}\\
\multicolumn{1}{c}{Add. Cond.:} &  \multicolumn{2}{c}{$B_z$}&  \multicolumn{2}{c|}{$|$\textbf{B}$|$}& \multicolumn{2}{c}{$B_z$}& \multicolumn{2}{c|}{$|$\textbf{B}$|$} & \multicolumn{2}{c}{$B_z$} & \multicolumn{2}{c|}{$|$\textbf{B}$|$} & \multicolumn{2}{c}{$B_z$} & \multicolumn{2}{c}{$|$\textbf{B}$|$}\\
\hline
				\multicolumn{1}{c|}{Level, Mm} & \colhead{IM} & \multicolumn{1}{c}{AS} & \colhead{IM} & \multicolumn{1}{c|}{AS}  & \colhead{IM} & \multicolumn{1}{c}{AS} & \colhead{IM} & \multicolumn{1}{c|}{AS} & \colhead{IM} & \multicolumn{1}{c}{AS} & \colhead{IM} & \multicolumn{1}{c|}{AS} & \colhead{IM} & \multicolumn{1}{c}{AS} & \colhead{IM} & \multicolumn{1}{c}{AS}
				}

\startdata
0.86 &     0.00 &     0.00 &     0.00 &     0.00 &     0.00 &     0.00
&     0.00 &     0.00 &     0.00 &     0.00 &     0.00 &     0.00 &
0.00 &     0.00 &
      0.00 &     0.00 \\
      1.29 &     5.08 &     5.90 &     4.11 & 5.68 &    10.00 &    11.72
&     9.60 &    11.70 &    14.51 &    16.32 &    13.65 &    16.20 &
5.61 &     6.35 &
      6.07 &     6.58 \\
      1.71 &     5.15 &     5.85 &     5.11 & 6.04 &     9.37 &    12.63
&     9.68 &    13.30 &    15.26 &    19.16 &    13.44 &    19.76 &
6.80 &     9.14 &
      8.92 &     9.58 \\
      2.14 &     4.21 &     4.54 &     0.00 & 1.50 &     9.01 &    11.13
&     8.57 &    11.08 &    15.12 &    16.77 &    13.52 &    16.85 &
0.00 &     1.84 &
      6.63 &     7.52 \\
      2.57 &     6.39 &     4.42 &     1.53 & 1.79 &    13.19 &    11.76
&     9.57 &    12.93 &    19.82 &    17.69 &    15.18 &    19.03 &
3.41 &     3.52 &
      8.07 &     7.82 \\
      3.00 &     6.02 &     4.55 &     2.76 & 2.36 &    13.59 &    13.25
&     9.93 &    14.62 &    20.75 &    19.41 &    16.09 &    19.71 &
5.63 &     4.97 &
      8.81 &     8.45 \\
      3.43 &     5.93 &     4.69 &     3.91 & 2.88 &    14.35 &    14.79
&    10.43 &    14.80 &    22.29 &    21.23 &    17.71 &    21.21 &
7.38 &     6.12 &
      9.91 &     9.01 \\
      3.86 &     6.15 &     4.95 &     5.05 & 3.38 &    15.14 &    16.18
&    10.96 &    15.60 &    25.19 &    23.69 &    20.38 &    23.71 &
8.68 &     7.29 &
     10.95 &     9.72 \\
      4.29 &     6.75 &     5.30 &     6.23 & 3.99 &    16.15 &    18.37
&    11.23 &    17.57 &    26.31 &    25.90 &    20.56 &    26.05 &
10.34 &     8.60 &
     12.32 &    10.79 \\
      4.71 &     7.65 &     5.77 &     7.48 & 4.57 &    17.99 &    21.39
&    11.86 &    20.27 &    28.44 &    29.42 &    22.33 &    29.79 &
12.41 &     9.97 &
     13.98 &    11.91 \\
      5.14 &     8.91 &     6.33 &     8.80 & 5.29 &    20.09 &    24.42
&    12.86 &    23.14 &    32.12 &    34.28 &    25.59 &    34.60 &
14.77 &    11.10 &
     15.78 &    12.95 \\
      5.57 &    10.47 &     6.98 &    10.17 & 6.01 &    22.32 &    27.50
&    14.02 &    26.07 &    36.48 &    39.06 &    29.29 &    39.06 &
17.43 &    12.38 &
     17.67 &    13.99 \\
      6.00 &    12.24 &     7.65 &    11.61 & 6.80 &    24.72 &    30.50
&    15.50 &    28.90 &    42.26 &    44.32 &    33.81 &    43.95 &
20.41 &    13.72 &
     19.68 &    15.18 \\
      6.43 &    14.18 &     8.27 &    13.09 & 7.49 &    27.16 &    33.35
&    17.13 &    31.68 &    49.27 &    50.21 &    39.13 &    49.77 &
23.71 &    15.12 &
     21.85 &    16.37 \\
      6.86 &    16.36 &     8.94 &    14.59 & 8.27 &    29.95 &    35.89
&    18.80 &    34.07 &    57.47 &    56.30 &    45.60 &    56.12 &
27.25 &    16.57 &
     24.14 &    17.60 \\
      7.29 &    18.82 &     9.67 &    16.08 & 9.04 &    33.34 &    38.34
&    20.47 &    36.33 &    66.57 &    62.11 &    53.37 &    62.06 &
30.98 &    17.91 &
     26.49 &    18.74 \\
      7.71 &    21.58 &    10.50 &    17.59 & 9.95 &    37.67 &    40.96
&    22.46 &    38.76 &    73.95 &    65.21 &    59.76 &    65.40 &
35.02 &    19.25 &
     28.94 &    19.91 \\
      8.14 &    24.72 &    11.43 &    19.15 & 10.88 &    43.56 &
44.26 &    25.18 &    41.91 &    79.86 &    66.46 &    64.80 &    66.81
&    39.40 &    20.63 &
     31.49 &    21.17 \\
      8.57 &    28.32 &    12.47 &    20.81 & 11.98 &    50.84 &
47.67 &    28.60 &    45.28 &    87.52 &    68.40 &    70.87 &    69.00
&    44.19 &    22.15 &
     34.14 &    22.60 \\
      9.00 &    32.47 &    13.69 &    22.63 & 13.20 &    58.84 &
50.91 &    32.12 &    48.59 &    97.54 &    71.09 &    78.41 &    71.80
&    49.69 &    23.99 &
     37.21 &    24.39 \\
      9.43 &    37.28 &    15.10 &    24.73 & 14.67 &    67.93 &
53.87 &    35.92 &    51.73 &   107.50 &    72.83 &    85.95 &    73.97
&    56.19 &    26.47 &
     40.97 &    26.83 \\
      9.86 &    42.97 &    16.72 &    27.26 & 16.36 &    77.80 &
57.06 &    40.00 &    55.20 &   117.33 &    73.25 &    93.28 &    74.90
&    64.00 &    29.31 &
     45.57 &    29.62 \\
     10.29 &    49.83 &    18.54 &    30.38 & 18.27 &    88.63 &
60.12 &    44.49 &    58.67 &   127.94 &    72.79 &   100.96 &    75.17
&    73.60 &    32.78 &
     51.28 &    33.08 \\
     10.71 &    58.21 &    20.55 &    34.31 & 20.43 &   101.09 &
63.55 &    49.38 &    62.65 &   138.33 &    69.66 &   108.55 &    72.77
&    85.71 &    36.83 &
     58.52 &    37.03 \\
     11.14 &    68.50 &    22.67 &    39.26 & 22.67 &   115.51 &
65.96 &    54.41 &    65.68 &   146.77 &    63.02 &   115.78 &    64.63
&   101.34 &    42.05 &
     68.01 &    42.19 \\
			\enddata
		\end{deluxetable}

\begin{deluxetable}{c|cccc|cccc|cccc|cccc}

\rotate
				
\tablecolumns{15}

\tablewidth{0pc}

\tabletypesize{\footnotesize}

\tablecaption{Normalized rms residual at a given level for bin-factor 9. \label{table_rms_res_2}}

\tablehead{
\multicolumn{1}{c|}{} & \multicolumn{4}{c|}{$\Delta_{rms}(B)$}& \multicolumn{4}{c|}{$\Delta_{rms}(B_x)$} & \multicolumn{4}{c|}{$\Delta_{rms}(B_y)$} & \multicolumn{4}{c}{$\Delta_{rms}(B_z)$}\\
\multicolumn{1}{c|}{Add. Cond.:} &  \multicolumn{2}{c}{$B_z$ \& $|$\textbf{B}$|$}&  \multicolumn{2}{c|}{\textbf{B}}& \multicolumn{2}{c}{$B_z$ \& $|$\textbf{B}$|$}& \multicolumn{2}{c|}{\textbf{B}} & \multicolumn{2}{c}{$B_z$ \& $|$\textbf{B}$|$} & \multicolumn{2}{c|}{\textbf{B}} & \multicolumn{2}{c}{$B_z$ \& $|$\textbf{B}$|$} & \multicolumn{2}{c}{\textbf{B}}\\
				\multicolumn{1}{c|}{Level, Mm} & \colhead{IM} & \multicolumn{1}{c}{AS} & \colhead{IM} & \multicolumn{1}{c|}{AS}  & \colhead{IM} & \multicolumn{1}{c}{AS} & \colhead{IM} & \multicolumn{1}{c|}{AS} & \colhead{IM} & \multicolumn{1}{c}{AS} & \colhead{IM} & \multicolumn{1}{c|}{AS} & \colhead{IM} & \multicolumn{1}{c}{AS} & \colhead{IM} & \multicolumn{1}{c}{AS}
				}
		%
%==========Table 3================================			
\startdata
0.86 &     0.00 &     0.00 &     0.00 &     0.00 &     0.00 &     0.00
&     0.00 &     0.00 &     0.00 &     0.00 &     0.00 &     0.00 &
0.00 &     0.00 &
      0.00 &     0.00 \\
      1.29 &     4.10 &     5.64 &     4.11 & 5.54 &     8.87 &    11.47
&     8.47 &    11.11 &    12.88 &    16.14 &    12.57 &    15.30 &
5.64 &     6.38 &
      5.52 &     6.19 \\
      1.71 &     5.20 &     5.95 &     5.28 & 5.71 &     8.96 &    12.80
&     7.80 &    11.28 &    12.27 &    19.31 &    10.99 &    17.24 &
6.81 &     9.34 &
      6.67 &     8.17 \\
      2.14 &     0.00 &     1.49 &     0.00 & 1.42 &     6.20 &     8.49
&     0.00 &     2.46 &    10.88 &    13.75 &     0.00 &     3.58 &
0.00 &     1.81 &
      0.00 &     1.64 \\
      2.57 &     1.12 &     1.73 &     0.97 & 1.46 &     7.22 &     8.90
&     3.29 &     4.30 &    11.51 &    14.40 &     6.10 &     7.91 &
3.03 &     3.45 &
      2.31 &     2.75 \\
      3.00 &     2.03 &     2.23 &     1.80 & 1.98 &     8.22 &    10.34
&     5.13 &     6.67 &    12.68 &    15.97 &     8.33 &    10.99 &
4.86 &     4.82 &
      4.21 &     4.23 \\
      3.43 &     2.90 &     2.77 &     2.60 & 2.38 &     9.35 &    11.95
&     6.57 &     9.88 &    14.65 &    18.09 &    11.06 &    14.92 &
6.16 &     6.00 &
      5.53 &     5.55 \\
      3.86 &     3.74 &     3.27 &     3.40 & 2.84 &    10.35 &    13.42
&     7.77 &    12.40 &    17.60 &    20.65 &    14.36 &    17.24 &
6.85 &     7.10 &
      6.29 &     6.42 \\
      4.29 &     4.58 &     3.79 &     4.19 & 3.33 &    10.85 &    15.38
&     8.49 &    14.07 &    17.60 &    22.99 &    14.43 &    19.91 &
7.75 &     8.40 &
      7.24 &     7.63 \\
      4.71 &     5.45 &     4.38 &     5.02 & 3.87 &    11.62 &    18.17
&     9.37 &    16.89 &    19.11 &    26.74 &    15.93 &    23.34 &
8.81 &     9.74 &
      8.36 &     8.99 \\
      5.14 &     6.36 &     5.02 &     5.89 & 4.50 &    12.74 &    20.83
&    10.59 &    19.68 &    21.56 &    31.58 &    18.07 &    28.61 &
9.82 &    10.90 &
      9.35 &    10.21 \\
      5.57 &     7.29 &     5.75 &     6.79 & 5.16 &    13.94 &    23.67
&    11.81 &    22.97 &    24.22 &    36.40 &    20.14 &    33.28 &
10.80 &    12.21 &
     10.28 &    11.54 \\
      6.00 &     8.28 &     6.51 &     7.74 & 5.85 &    15.30 &    26.33
&    13.25 &    25.85 &    27.26 &    41.72 &    22.85 &    38.88 &
11.80 &    13.69 &
     11.25 &    13.05 \\
      6.43 &     9.31 &     7.27 &     8.74 & 6.47 &    16.81 &    29.06
&    14.87 &    29.07 &    30.69 &    48.28 &    26.17 &    44.58 &
12.84 &    15.18 &
     12.29 &    14.63 \\
      6.86 &    10.32 &     8.08 &     9.72 & 7.18 &    18.40 &    31.40
&    16.50 &    31.66 &    34.91 &    55.15 &    30.18 &    50.36 &
13.81 &    16.75 &
     13.33 &    16.29 \\
      7.29 &    11.25 &     8.94 &    10.64 & 7.87 &    19.89 &    33.75
&    17.98 &    34.44 &    40.38 &    61.97 &    35.39 &    55.99 &
14.56 &    18.12 &
     14.25 &    17.84 \\
      7.71 &    12.09 &     9.89 &    11.49 & 8.73 &    21.46 &    36.28
&    19.53 &    37.06 &    44.52 &    65.93 &    39.08 &    59.39 &
15.16 &    19.58 &
     15.10 &    19.40 \\
      8.14 &    12.82 &    10.89 &    12.27 & 9.62 &    23.31 &    39.65
&    21.32 &    40.57 &    47.26 &    68.15 &    41.36 &    60.75 &
15.53 &    20.95 &
     15.72 &    20.92 \\
      8.57 &    13.43 &    11.99 &    12.93 & 10.77 &    25.57 &
43.24 &    23.45 &    44.06 &    50.42 &    70.73 &    44.37 &    63.01
&    15.55 &    22.55 &
     15.97 &    22.47 \\
      9.00 &    13.93 &    13.24 &    13.49 & 11.98 &    27.82 &
46.96 &    25.48 &    47.56 &    53.95 &    74.24 &    47.73 &    65.69
&    15.66 &    24.33 &
     16.24 &    24.26 \\
      9.43 &    14.43 &    14.71 &    14.04 & 13.62 &    30.04 &
50.52 &    27.44 &    50.70 &    57.27 &    76.45 &    50.95 &    68.00
&    16.20 &    26.83 &
     16.86 &    26.69 \\
      9.86 &    14.98 &    16.42 &    14.60 & 15.36 &    32.19 &
54.43 &    29.27 &    54.29 &    59.95 &    77.38 &    53.68 &    69.17
&    17.12 &    29.59 &
     17.76 &    29.44 \\
     10.29 &    15.55 &    18.34 &    15.15 & 17.51 &    34.22 &
58.26 &    30.99 &    57.79 &    62.06 &    76.82 &    56.10 &    69.56
&    18.33 &    33.15 &
     18.91 &    32.88 \\
     10.71 &    16.21 &    20.51 &    15.70 & 19.82 &    35.85 &
62.49 &    32.34 &    61.96 &    64.01 &    72.92 &    58.56 &    68.52
&    19.96 &    37.11 &
     20.43 &    36.86 \\
     11.14 &    17.11 &    22.72 &    16.37 & 22.44 &    36.79 &
65.66 &    33.11 &    65.55 &    66.52 &    63.97 &    61.44 &    63.54
&    22.38 &    42.33 &
     22.62 &    42.05 \\
\enddata
		\end{deluxetable}

\begin{deluxetable}{c|cccc|cccc|cccc|cccc}

\rotate
				
\tablecolumns{15}

\tablewidth{0pc}

\tabletypesize{\footnotesize}

\tablecaption{Normalized rms residual at a given level for bin-factor 9 for the location 30$^\circ$ North at the central meridian. \label{table_rms_res_3}}

\tablehead{
\multicolumn{1}{c|}{} & \multicolumn{4}{c|}{$\Delta_{rms}(B)$}& \multicolumn{4}{c|}{$\Delta_{rms}(B_x)$} & \multicolumn{4}{c|}{$\Delta_{rms}(B_y)$} & \multicolumn{4}{c}{$\Delta_{rms}(B_z)$}\\
\multicolumn{1}{c|}{Add. Cond.:} &  \multicolumn{2}{c}{$B_{||}$}&  \multicolumn{2}{c|}{$B_{||}$ \& $|$\textbf{B}$|$}& \multicolumn{2}{c}{$B_{||}$}& \multicolumn{2}{c|}{$B_{||}$ \& $|$\textbf{B}$|$} & \multicolumn{2}{c}{$B_{||}$} & \multicolumn{2}{c|}{$B_{||}$ \& $|$\textbf{B}$|$} & \multicolumn{2}{c}{$B_{||}$} & \multicolumn{2}{c}{$B_{||}$ \& $|$\textbf{B}$|$}\\
				\multicolumn{1}{c|}{Level, Mm} & \colhead{IM} & \multicolumn{1}{c}{AS} & \colhead{IM} & \multicolumn{1}{c|}{AS}  & \colhead{IM} & \multicolumn{1}{c}{AS} & \colhead{IM} & \multicolumn{1}{c|}{AS} & \colhead{IM} & \multicolumn{1}{c}{AS} & \colhead{IM} & \multicolumn{1}{c|}{AS} & \colhead{IM} & \multicolumn{1}{c}{AS} & \colhead{IM} & \multicolumn{1}{c}{AS}
				}
		%
%==========================Table 4================================			
\startdata
% new			
     0.86 &     0.00 &     0.00 &     0.00 &     0.00 &     0.00 &
0.00 &     0.00 &     0.00 &     0.00 &     0.00 &     0.00 &     0.00
&     0.00 &     0.00 &
      0.00 &     0.00 \\
      1.29 &     4.95 &     5.90 &     4.09 & 5.64 &    10.31 &    11.78
&     9.01 &    11.54 &    13.97 &    16.38 &    13.10 &    16.15 &
5.65 &     6.38 &
      5.79 &     6.40 \\
      1.71 &     4.68 &     5.89 &     5.20 & 5.94 &    10.20 &    12.95
&     8.84 &    13.05 &    13.63 &    19.26 &    12.34 &    19.42 &
6.60 &     9.29 &
      7.47 &     9.39 \\
      2.14 &     3.73 &     4.66 &     0.00 & 1.53 &     9.96 &    11.86
&     6.63 &     9.56 &    11.55 &    17.54 &     9.91 &    14.39 &
3.78 &     4.57 &
      3.24 &     3.68 \\
      2.57 &     6.06 &     4.55 &     1.22 & 1.77 &    14.09 &    12.72
&     7.65 &    10.17 &    16.25 &    18.99 &    11.29 &    15.41 &
5.50 &     5.55 &
      4.61 &     4.74 \\
      3.00 &     5.55 &     4.65 &     2.21 & 2.28 &    14.20 &    14.12
&     8.47 &    11.65 &    17.27 &    20.62 &    12.37 &    16.89 &
6.78 &     6.71 &
      6.03 &     5.95 \\
      3.43 &     5.38 &     4.83 &     3.15 & 2.79 &    14.73 &    15.40
&     9.36 &    12.95 &    18.95 &    22.13 &    14.43 &    18.55 &
8.13 &     7.69 &
      7.24 &     7.00 \\
      3.86 &     5.59 &     5.10 &     4.08 & 3.29 &    15.26 &    16.54
&    10.18 &    14.09 &    21.68 &    24.45 &    17.56 &    20.89 &
9.37 &     8.60 &
      8.08 &     7.94 \\
      4.29 &     6.24 &     5.51 &     5.04 & 3.83 &    15.86 &    18.51
&    10.53 &    15.84 &    21.43 &    26.72 &    16.93 &    23.03 &
11.04 &     9.76 &
      9.17 &     9.14 \\
      4.71 &     7.26 &     6.01 &     6.07 & 4.42 &    17.17 &    21.42
&    11.14 &    18.49 &    22.48 &    30.09 &    17.17 &    26.61 &
13.07 &    11.10 &
     10.51 &    10.50 \\
      5.14 &     8.60 &     6.65 &     7.17 & 5.10 &    18.75 &    24.32
&    12.15 &    21.05 &    25.37 &    34.91 &    18.64 &    31.22 &
15.30 &    12.30 &
     11.88 &    11.73 \\
      5.57 &    10.20 &     7.36 &     8.29 & 5.87 &    20.55 &    27.42
&    13.31 &    23.93 &    29.00 &    39.65 &    20.81 &    36.07 &
17.72 &    13.58 &
     13.24 &    13.06 \\
      6.00 &    11.95 &     8.11 &     9.49 & 6.66 &    22.53 &    30.37
&    14.72 &    26.58 &    33.87 &    44.91 &    23.67 &    41.35 &
20.38 &    14.91 &
     14.62 &    14.53 \\
      6.43 &    13.78 &     8.79 &    10.71 & 7.45 &    24.54 &    33.26
&    16.29 &    29.30 &    39.74 &    50.97 &    26.98 &    48.03 &
23.34 &    16.27 &
     16.08 &    15.98 \\
      6.86 &    15.78 &     9.52 &    11.93 & 8.27 &    26.87 &    35.76
&    17.92 &    31.63 &    46.81 &    57.37 &    31.44 &    55.03 &
26.55 &    17.64 &
     17.56 &    17.51 \\
      7.29 &    17.95 &    10.26 &    13.11 & 9.13 &    29.82 &    38.23
&    19.46 &    33.96 &    55.21 &    63.30 &    37.43 &    61.98 &
29.94 &    18.89 &
     18.99 &    18.80 \\
      7.71 &    20.33 &    11.11 &    14.28 & 10.06 &    33.66 &
40.80 &    21.11 &    36.43 &    62.42 &    66.29 &    41.94 &    65.92
& 33.55 &    20.15 &
     20.45 &    20.18 \\
      8.14 &    23.01 &    11.97 &    15.48 & 11.03 &    38.83 &
44.11 &    23.13 &    39.74 &    68.38 &    67.32 &    45.23 &    67.98
& 37.34 &    21.48 &
     21.86 &    21.48 \\
      8.57 &    26.04 &    12.97 &    16.71 & 12.08 &    45.00 &
47.51 &    25.66 &    43.27 &    75.97 &    69.15 &    49.42 &    70.36
& 41.35 &    23.00 &
     23.16 &    23.05 \\
      9.00 &    29.49 &    14.08 &    18.02 & 13.29 &    51.32 &
50.78 &    28.25 &    46.91 &    86.00 &    71.78 &    54.11 &    73.60
& 45.88 &    24.82 &
     24.69 &    24.80 \\
      9.43 &    33.47 &    15.43 &    19.51 & 14.72 &    58.11 &
53.77 &    30.97 &    50.40 &    96.78 &    73.52 &    58.49 &    75.74
& 51.25 &    27.31 &
     26.70 &    27.31 \\
      9.86 &    38.17 &    16.96 &    21.25 & 16.40 &    65.27 &
56.98 &    33.86 &    54.27 &   108.33 &    73.88 &    62.59 &    76.70
& 57.74 &    30.07 &
     29.24 &    30.02 \\
     10.29 &    43.88 &    18.74 &    23.30 & 18.32 &    72.92 &
60.05 &    36.99 &    58.06 &   121.40 &    73.40 &    66.98 &    76.51
& 65.79 &    33.55 &
     32.44 &    33.57 \\
     10.71 &    50.90 &    20.68 &    25.73 & 20.50 &    81.17 &
63.50 &    40.20 &    62.34 &   135.36 &    70.01 &    72.03 &    73.15
& 76.07 &    37.39 &
     36.62 &    37.42 \\
     11.14 &    59.56 &    22.74 &    28.67 & 22.73 &    89.16 &
65.93 &    43.35 &    65.58 &   151.42 &    63.14 &    78.63 &    64.34
& 89.60 &    42.54 &
     42.35 &    42.59 \\
    			\enddata
		\end{deluxetable}

\begin{deluxetable}{c|cccc|cccc|cccc|cccc}

\rotate
				
\tablecolumns{15}

\tablewidth{0pc}

\tabletypesize{\footnotesize}

\tablecaption{Normalized rms residual at a given level for bin-factor 9 for the location 30$^\circ$ North and 30$^\circ$ West. \label{table_rms_res_4}}

\tablehead{
\multicolumn{1}{c|}{} & \multicolumn{4}{c|}{$\Delta_{rms}(B)$}& \multicolumn{4}{c|}{$\Delta_{rms}(B_x)$} & \multicolumn{4}{c|}{$\Delta_{rms}(B_y)$} & \multicolumn{4}{c}{$\Delta_{rms}(B_z)$}\\
\multicolumn{1}{c|}{Add. Cond.:} &  \multicolumn{2}{c}{$B_{||}$}&  \multicolumn{2}{c|}{$B_{||}$ \& $|$\textbf{B}$|$}& \multicolumn{2}{c}{$B_{||}$}& \multicolumn{2}{c|}{$B_{||}$ \& $|$\textbf{B}$|$} & \multicolumn{2}{c}{$B_{||}$} & \multicolumn{2}{c|}{$B_{||}$ \& $|$\textbf{B}$|$} & \multicolumn{2}{c}{$B_{||}$} & \multicolumn{2}{c}{$B_{||}$ \& $|$\textbf{B}$|$}\\
				\multicolumn{1}{c|}{Level, Mm} & \colhead{IM} & \multicolumn{1}{c}{AS} & \colhead{IM} & \multicolumn{1}{c|}{AS}  & \colhead{IM} & \multicolumn{1}{c}{AS} & \colhead{IM} & \multicolumn{1}{c|}{AS} & \colhead{IM} & \multicolumn{1}{c}{AS} & \colhead{IM} & \multicolumn{1}{c|}{AS} & \colhead{IM} & \multicolumn{1}{c}{AS} & \colhead{IM} & \multicolumn{1}{c}{AS}
				}
		%
%======================Table 5============================			
\startdata
% new			
      0.86 &     0.00 &     0.00 &     0.00 & 0.00 &     0.00 &     0.00
&     0.00 &     0.00 &     0.00 &     0.00 &     0.00 &     0.00 &
0.00 &     0.00 &
      0.00 &     0.00 \\
      1.29 &     4.94 &     5.93 &     4.09 & 5.64 &     9.99 &    11.78
&     8.92 &    11.54 &    14.28 &    16.55 &    13.14 &    16.16 &
5.64 &     6.43 &
      5.83 &     6.43 \\
      1.71 &     4.45 &     5.93 &     5.19 & 5.94 &     9.48 &    13.21
&     8.64 &    13.16 &    14.73 &    20.10 &    11.67 &    19.86 &
6.34 &     9.43 &
      7.67 &     9.39 \\
      2.14 &     3.52 &     5.00 &     0.00 & 1.65 &     7.90 &    12.54
&     5.27 &     9.87 &    12.50 &    18.67 &     9.03 &    15.53 &
4.62 &     6.51 &
      3.66 &     5.50 \\
      2.57 &     6.11 &     4.90 &     1.26 & 1.91 &    11.73 &    13.48
&     6.18 &    10.57 &    18.05 &    20.34 &    11.15 &    16.98 &
6.31 &     6.63 &
      4.96 &     5.74 \\
      3.00 &     5.57 &     5.00 &     2.31 & 2.46 &    12.23 &    14.79
&     7.09 &    12.05 &    18.60 &    21.93 &    12.16 &    18.59 &
7.10 &     7.28 &
      6.24 &     6.47 \\
      3.43 &     5.46 &     5.19 &     3.32 & 3.01 &    13.04 &    15.90
&     8.11 &    13.31 &    20.11 &    23.24 &    14.21 &    19.97 &
8.22 &     7.99 &
      7.44 &     7.15 \\
      3.86 &     5.81 &     5.46 &     4.34 & 3.54 &    13.92 &    16.96
&     9.06 &    14.37 &    22.94 &    25.64 &    17.21 &    22.07 &
9.42 &     8.70 &
      8.34 &     7.95 \\
      4.29 &     6.63 &     5.83 &     5.40 & 4.05 &    15.08 &    18.93
&     9.64 &    16.17 &    23.01 &    28.38 &    16.53 &    24.54 &
11.14 &     9.82 &
      9.47 &     9.16 \\
      4.71 &     7.81 &     6.28 &     6.53 & 4.62 &    17.13 &    21.89
&    10.67 &    18.71 &    24.35 &    32.21 &    17.20 &    28.40 &
13.30 &    11.06 &
     10.83 &    10.56 \\
      5.14 &     9.30 &     6.84 &     7.72 & 5.24 &    19.53 &    24.80
&    12.14 &    21.29 &    27.82 &    36.98 &    19.40 &    33.28 &
15.75 &    12.25 &
     12.26 &    11.85 \\
      5.57 &    11.03 &     7.48 &     8.94 & 5.95 &    22.12 &    27.98
&    13.66 &    24.16 &    32.36 &    41.39 &    21.94 &    37.83 &
18.45 &    13.38 &
     13.70 &    13.12 \\
      6.00 &    12.93 &     8.14 &    10.24 & 6.66 &    24.98 &    31.02
&    15.39 &    26.90 &    38.62 &    46.45 &    25.02 &    43.22 &
21.42 &    14.53 &
     15.19 &    14.47 \\
      6.43 &    14.97 &     8.72 &    11.57 & 7.36 &    27.94 &    33.99
&    17.22 &    29.64 &    46.04 &    52.02 &    28.22 &    49.47 &
24.67 &    15.68 &
     16.79 &    15.78 \\
      6.86 &    17.25 &     9.37 &    12.91 & 8.08 &    31.23 &    36.50
&    19.03 &    32.00 &    54.35 &    57.90 &    31.96 &    56.25 &
28.13 &    16.89 &
     18.46 &    17.12 \\
      7.29 &    19.83 &    10.04 &    14.21 & 8.85 &    35.12 &    38.95
&    20.73 &    34.29 &    63.70 &    63.08 &    37.21 &    62.66 &
31.73 &    18.04 &
     20.10 &    18.28 \\
      7.71 &    22.75 &    10.84 &    15.48 & 9.70 &    39.73 &    41.42
&    22.64 &    36.68 &    71.91 &    65.44 &    41.46 &    66.13 &
35.50 &    19.28 &
     21.73 &    19.52 \\
      8.14 &    26.11 &    11.70 &    16.73 & 10.63 &    45.36 &
44.66 &    25.12 &    39.89 &    79.07 &    65.90 &    44.62 &    67.40
& 39.49 &    20.63 &
     23.29 &    20.78 \\
      8.57 &    29.99 &    12.71 &    17.96 & 11.69 &    51.81 &
47.92 &    28.18 &    43.28 &    88.00 &    67.41 &    48.63 &    69.24
& 43.84 &    22.20 &
     24.72 &    22.29 \\
      9.00 &    34.48 &    13.87 &    19.21 & 12.91 &    58.50 &
51.05 &    31.25 &    46.79 &    99.29 &    69.75 &    53.36 &    71.89
& 48.93 &    24.10 &
     26.39 &    24.09 \\
      9.43 &    39.68 &    15.26 &    20.55 & 14.37 &    66.08 &
53.84 &    34.52 &    50.08 &   111.10 &    71.62 &    58.39 &    73.93
& 55.13 &    26.65 &
     28.58 &    26.61 \\
      9.86 &    45.81 &    16.84 &    22.10 & 16.11 &    74.87 &
56.93 &    38.04 &    53.87 &   123.31 &    72.35 &    63.32 &    74.95
& 62.77 &    29.53 &
     31.33 &    29.45 \\
     10.29 &    53.18 &    18.65 &    23.91 & 18.06 &    85.43 &
59.88 &    41.91 &    57.58 &   136.54 &    72.49 &    68.41 &    75.16
& 72.38 &    33.04 &
     34.80 &    33.01 \\
     10.71 &    62.04 &    20.62 &    26.15 & 20.32 &    98.41 &
63.33 &    46.09 &    61.98 &   149.67 &    70.13 &    73.92 &    73.09
& 84.68 &    37.02 &
     39.30 &    37.02 \\
     11.14 &    72.72 &    22.73 &    29.02 & 22.64 &   113.43 &
65.88 &    50.42 &    65.48 &   162.37 &    63.51 &    80.21 &    64.75
& 100.80 &    42.20 &
     45.43 &    42.21 \\
    		\enddata
		\end{deluxetable}

\clearpage

\begin{deluxetable}{c|ccc|ccc|ccc|ccc}

\rotate
				
\tablecolumns{15}

\tablewidth{0pc}

\tabletypesize{\footnotesize}

\tablecaption{Normalized rms residual at a given level for bin-factor 9 for reconstructions with the IM code from the chromospheric boundary obtained by the AS code. \label{table_rms_res_5}}

\tablehead{
\multicolumn{1}{c|}{} & \multicolumn{3}{c|}{$\Delta_{rms}(B)$}& \multicolumn{3}{c|}{$\Delta_{rms}(Bx)$} & \multicolumn{3}{c|}{$\Delta_{rms}(By)$} & \multicolumn{3}{c}{$\Delta_{rms}(Bz)$}\\
\multicolumn{1}{c|}{Level, Mm} &  \multicolumn{1}{c}{$B_{||}$} & \multicolumn{1}{c}{$|$\textbf{B}$|$} & \multicolumn{1}{c|}{$B_{||}$ \& $|$\textbf{B}$|$} & \multicolumn{1}{c}{$B_{||}$} & \multicolumn{1}{c}{$|$\textbf{B}$|$} & \multicolumn{1}{c|}{$B_{||}$ \& $|$\textbf{B}$|$} & \multicolumn{1}{c}{$B_{||}$} & \multicolumn{1}{c}{$|$\textbf{B}$|$} & \multicolumn{1}{c|}{$B_{||}$ \& $|$\textbf{B}$|$} & \multicolumn{1}{c}{$B_{||}$} & \multicolumn{1}{c}{$|$\textbf{B}$|$} & \multicolumn{1}{c}{$B_{||}$ \& $|$\textbf{B}$|$}\\
}
	
%========================Table 6=========================		
\startdata
%	new		
     0.86 &     0.00 &     0.00 &     0.00 & 0.00 &     0.00 &     0.00
&     0.00 &     0.00 &     0.00 &     0.00 &     0.00 &     0.00 \\
      1.29 &     5.90 &     5.68 &     5.64 & 11.72 &    11.70 &
11.47 &    16.32 &    16.20 &    16.14 &     6.35 &     6.58 &     6.38 \\
      1.71 &     5.85 &     6.04 &     5.95 & 12.63 &    13.30 &
12.80 &    19.16 &    19.76 &    19.31 &     9.14 &     9.58 &     9.34 \\
      2.14 &     4.54 &     1.50 &     1.49 & 11.13 &    11.08 &
8.49 &    16.77 &    16.85 &    13.75 &     1.84 &     7.52 &     1.81 \\
      2.57 &     4.23 &     1.64 &     1.59 & 10.75 &    10.79 &
7.56 &    17.14 &    16.97 &    13.63 &     3.37 &     6.91 &     3.33 \\
      3.00 &     4.34 &     2.18 &     2.13 & 11.32 &    10.26 &
7.89 &    18.30 &    17.20 &    14.32 &     5.16 &     7.42 &     5.07 \\
      3.43 &     4.57 &     2.73 &     2.73 & 12.07 &    10.21 &
8.68 &    20.22 &    18.08 &    15.97 &     6.57 &     8.42 &     6.49 \\
      3.86 &     4.90 &     3.27 &     3.38 & 12.95 &    10.61 &
9.57 &    23.27 &    20.41 &    18.60 &     7.52 &     9.32 &     7.46 \\
      4.29 &     5.34 &     3.83 &     4.05 & 14.32 &    11.09 &
10.33 &    24.76 &    20.97 &    19.03 &     8.61 &    10.42 &     8.62 \\
      4.71 &     5.88 &     4.43 &     4.79 & 16.35 &    11.93 &
11.46 &    27.59 &    23.16 &    21.05 &     9.78 &    11.59 &     9.85 \\
      5.14 &     6.51 &     5.09 &     5.59 & 18.49 &    12.95 &
12.85 &    31.79 &    26.56 &    24.15 &    10.89 &    12.68 &    11.00 \\
      5.57 &     7.21 &     5.77 &     6.44 & 20.46 &    13.90 &
14.22 &    36.29 &    29.92 &    27.23 &    12.03 &    13.75 &    12.13 \\
      6.00 &     7.97 &     6.57 &     7.37 & 22.45 &    15.01 &
15.74 &    41.53 &    33.87 &    30.89 &    13.23 &    14.81 &    13.26 \\
      6.43 &     8.76 &     7.43 &     8.35 & 24.37 &    16.23 &
17.36 &    47.46 &    38.52 &    35.15 &    14.55 &    15.93 &    14.45 \\
      6.86 &     9.63 &     8.36 &     9.35 & 26.20 &    17.45 &
19.01 &    54.18 &    44.08 &    40.18 &    15.91 &    17.10 &    15.63 \\
      7.29 &    10.55 &     9.31 &    10.30 & 27.91 &    18.55 &
20.55 &    61.18 &    50.33 &    46.06 &    17.18 &    18.23 &    16.71 \\
      7.71 &    11.53 &    10.28 &    11.20 & 29.79 &    19.75 &
22.18 &    65.71 &    54.52 &    50.04 &    18.41 &    19.37 &    17.72 \\
      8.14 &    12.56 &    11.24 &    12.04 & 32.24 &    21.37 &
24.21 &    68.48 &    57.16 &    52.46 &    19.53 &    20.45 &    18.55 \\
      8.57 &    13.66 &    12.19 &    12.80 & 35.07 &    23.50 &
26.63 &    72.12 &    60.48 &    55.55 &    20.44 &    21.35 &    19.10 \\
      9.00 &    14.89 &    13.18 &    13.50 & 37.63 &    25.57 &
28.91 &    76.86 &    64.45 &    59.21 &    21.51 &    22.43 &    19.72 \\
      9.43 &    16.30 &    14.27 &    14.22 & 40.10 &    27.65 &
31.14 &    81.23 &    67.89 &    62.48 &    22.96 &    23.92 &    20.65 \\
      9.86 &    17.97 &    15.50 &    14.99 & 42.43 &    29.72 &
33.24 &    85.11 &    70.40 &    64.97 &    24.79 &    25.78 &    21.85 \\
     10.29 &    19.92 &    16.84 &    15.80 & 44.67 &    31.75 &
35.17 &    88.96 &    72.14 &    66.94 &    27.02 &    28.03 &    23.28 \\
     10.71 &    22.31 &    18.38 &    16.73 & 46.67 &    33.50 &
36.63 &    92.54 &    72.99 &    68.52 &    29.85 &    30.82 &    25.06 \\
     11.14 &    25.28 &    20.24 &    17.88 & 48.20 &    34.67 &
37.30 &    95.87 &    73.05 &    70.03 &    33.69 &    34.45 &    27.48 \\
\enddata
		\end{deluxetable}

\end{document}